\documentclass[12pt]{article}
\pdfoutput=1
\usepackage[square,comma,numbers,sort&compress]{natbib}
\usepackage{graphicx,epstopdf,amssymb,amsfonts,amsmath,amsthm,array,
mathrsfs,amscd}
\usepackage{float}
\usepackage{xcolor}
\usepackage{appendix}
\usepackage{hyperref}
\newcommand{\pbs}[1]{\let\temp=\\#1\let\\=\temp}
\numberwithin{equation}{section}
\def\be{\begin{equation}}\def\ee{\end{equation}}
\def\cvp{\raise 2pt\hbox{,}} 
 \def\tr{\mathop{\text{tr}}\nolimits}

\def\re{\mathop{\text{Re}}\nolimits}

\def\la{\lambda}
 \def\uN{\text{U}(N)}

\def\gG{\mathcal G}\def\gB{\mathcal B}\def\gJ{\mathcal J}
\def\gK{\mathcal K}
\def\gT{\mathcal T}
\def\gM{\mathcal M}

\newtheorem{lemma}{Lemma}[section]
\newtheorem{proposition}{Proposition}[section]
\newtheorem{theorem}{Theorem}[section]
\DeclareMathOperator{\ind}{ind}

\def\npb#1#2#3{{\it Nucl.\ Phys.\ }{\bf B #1} (#2) #3}
\def\npps#1#2#3{{\it Nucl.\ Phys.\ Proc.\ Suppl.\ }{\bf #1} (#2) #3}

\def\jhep#1#2#3{{\it J. High Energy Phys.\ }{\bf #1} (#2) #3}
\def\prd#1#2#3{{\it Phys.\ Rev.\ }{\bf D #1} (#2) #3}

\def\prx#1#2#3{{\it Phys.\ Rev.\ }{\bf X #1} (#2) #3}

\def\cmp#1#2#3{{\it Comm.\ Math.\ Phys.\ }{\bf #1} (#2) #3}
\def\pr#1#2#3{{\it Phys.\ Rep.\ }{\bf #1} (#2) #3}
\def\jmp#1#2#3{{\it J.\ Math.\ Phys.\ }{\bf #1} (#2) #3}

\def\mpla#1#2#3{{\it Mod.\ Phys.\ Lett.\ }{\bf A #1} (#2) #3}

\def\imath#1#2#3{{\it Invent math }{\bf #1} (#2) #3}

\begin{document}
%
%
{\pagestyle{empty}
\parskip 0in
\

\vfill
\begin{center}
{\LARGE A New Large $N$ Expansion}

\bigskip

{\LARGE for General Matrix-Tensor Models}

\vspace{0.4in}


Frank F{\scshape errari},${}^{1,2}$ Vincent R{\scshape ivasseau}${}^{3}$ and Guillaume~V{\scshape alette}${}^{1}$ 
\\

\medskip
${}^1${\it Service de Physique Th\'eorique et Math\'ematique\\
Universit\'e Libre de Bruxelles (ULB) and International Solvay Institutes\\
Campus de la Plaine, CP 231, B-1050 Bruxelles, Belgique}

\smallskip

${}^{2}${\it Fields, Gravity and Strings\\
Center for the Theoretical Physics of the Universe\\
Institute for Basic Sciences, Daejeon, 34047 South Korea}

\smallskip

${}^{3}${\it Laboratoire de Physique Th\'eorique, CNRS UMR 8627, Universit\'e Paris-Sud, 91405 Orsay Cedex, France}


\smallskip
{\tt frank.ferrari@ulb.ac.be, vincent.rivasseau@th.u-psud.fr, guillaume.valette@ulb.ac.be}
\end{center}
\vfill\noindent

We define a new large $N$ limit for general $\text{O}(N)^{R}$ or $\text{U}(N)^{R}$ invariant tensor models, based on an enhanced large $N$ scaling of the coupling constants. The resulting large $N$ expansion is organized in terms of a half-integer associated with Feynman graphs that we call the index. This index has a natural interpretation in terms of the many matrix models embedded in the tensor model. Our new scaling can be shown to be optimal for a wide class of non-melonic interactions, which includes all the maximally single-trace terms. Our construction allows to define a new large $D$ expansion of the sum over diagrams of fixed genus in matrix models with an additional $\text{O}(D)^{r}$ global symmetry. When the interaction is the complete vertex of order $R+1$, we identify in detail the leading order graphs for $R$ a prime number. This slightly surprising condition is equivalent to the complete interaction being maximally single-trace.

\vfill

\medskip
%
%
\newpage\pagestyle{plain}
\baselineskip 16pt
\setcounter{footnote}{0}

}

\tableofcontents

\section{\label{s1Sec}Introduction}

\subsection{Generalities}

The modern theory of random tensors \cite{Guraubook} relies on the discovery 
of the tensor $1/N$ expansion \cite{1/N}, which generalizes 
the standard topological expansion of matrix models \cite{tHooftplanar}. The initial version is
simply called the ``colored model"  \cite{colored}. It is made of $R+1$ different complex tensors of rank $R$,  $T^{j}_{a_{1}\cdots a_{R}}$, with $j =0, \ldots, R$,
$a_i = 1, \ldots , N$ for $i= 1, \ldots , R$.
The symmetry group is $\text{U}(N)^{R(R+1)/2}$ and the canonical interaction is
obtained by contracting the indices of these $R+1$ tensors (and separately for their complex conjugates) according to the pattern of the complete graph $K_{R+1}$ on $R+1$ vertices. The tensor $1/N$ expansion is indexed by a new integer, the \emph{Gurau degree}, hereafter simply called degree.
It generalizes the two-dimensional genus, but is no longer a topological invariant for $R \ge 3$. 
Leading Feynman graphs have degree zero. They are called melons and their structure
was identified in \cite{melons}. Surprisingly, melons are more restricted than planar graphs; they could also be called ``super planar'' since there are $\frac{1}{2}R!$ different canonical ways to draw them on a plane, associated to so-called ``jackets," as we shall review below.

The vacuum Feynman graphs of the colored model are $(R+1)$-regular edge-colored bipartite graphs, a category dual to the colored
triangulations of orientable piecewise linear quasi-manifolds in dimension $R$ \cite{geom}. It is well-known that tensor amplitudes
ponder the corresponding dual spaces with a discretized version of the Einstein-Hilbert
action \cite{ambjorn}. The $1/N$ random tensor expansion, by introducing a hierarchy in this pondered 
sum over random geometries, opens a promising new perspective on quantum gravity  in arbitrary dimension, 
nicknamed the \emph{tensor track} \cite{track}. It generalizes the older relationship between random matrices,
random geometry and  quantum gravity in two dimensions \cite{DiFrancesco:1993nw}.

``Uncoloring" \cite{univ,BGR} is a way to vastly generalize the $\text{U}(N)^{R(R+1)/2}$ invariant colored theory to $\text{U}(N)^{R}$ invariant models made of tensors of rank $R$, where $\text{U}(N)^{R}$ corresponds to invariance
under \emph{independent} unitary changes of basis for \emph{each index} of the tensor. Any $\text{U}(N)^{R}$ invariant interaction vertex can be a priori considered. For $R\geq 3$, this yields a much richer class of possibilities than in the vector ($R=1$) or matrix ($R=2$) cases. Actually, the possible \emph{interactions} and \emph{vacuum Feynman graphs} of the uncolored models at rank $R$ coincide with the possible \emph{vacuum Feynman graphs} of the colored theory at rank $R-1$ and $R$, respectively. To define the $1/N$ expansion of the uncolored models, BGR introduced in \cite{BGR} a particular  scaling of the coupling constants at large $N$ (the analogue of 't~Hooft's scaling for matrix models) according to the degree of the associated interaction viewed as a vacuum graph of the rank $R-1$ underlying colored theory. This yields a well-defined $1/N$ expansion, which, remarkably, is itself indexed by the degree of the Feynman graphs.

We say that a large $N$ scaling for a particular coupling constant is \emph{optimal} if it is impossible to enhance the scaling and still have a well-defined $1/N$ expansion. For example, if an interaction vertex can appear an arbitrary number of times in Feynman graphs at any fixed order in the $1/N$ expansion, then the scaling for this vertex is necessarily optimal, since any further enhancement would produce diagrams proportional to an arbitrarily high power of $N$. It is straightforward to see that the BGR scaling is optimal for \emph{melonic} interactions, but is not optimal in general. For a particular interaction, the optimal scaling, if it exists at all, can be very complicated to compute. Optimal scalings are understood only for a very small subset of all possible interactions \cite{Bonzom1bis}. A striking example of non-BGR scalings is given in \cite{BDR}, where a mixture of melonic and non-melonic quartic interactions at rank four is used to interpolate between the usual tensor expansion 
dominated by melonic graphs and the ordinary topological expansion of matrix models dominated by planar graphs.
A non-trivial phase of ``baby universes" occurs at the transition point. Other examples treated in the literature
include meander, octaedric and order 6 interactions up to rank 4 \cite{Bonzom2,LionniTh}, and the most general results up to now are summarized in \cite{Bonzometal}.

The analysis of \cite{BGR} can be generalized to more general symmetry groups. For instance a real tensor model has $\text{O}(N)^{R}$
symmetry and its tensor space is obtained by removing the bipartite/orientability condition (see below). 
In \cite{CarrozzaTanasa}, a particular tensor model in rank three with such an $\text{O}(N)^{3}$ symmetry  is introduced
and studied. It uses two different quartic interactions. One is ``tetraedric," based on a three-regular edge-coloring of the complete graph $K_4$, and the other is melonic. Interestingly, the authors do not use the BGR scaling for the large $N$ limit, 
since it would wash away the non-melonic tetraedric interaction that they want to preserve. They enhance the tetraedric scaling so that, at a given order in the large $N$ expansion, many more Feynman diagrams can contribute than with the BGR scaling. 
The remarkable point of \cite{CarrozzaTanasa} is that, in spite of this non-trivial enhancement, 
the large $N$ limit still exists and the leading graphs can be identified. In particular, they are no longer melonic in the traditional BGR sense. However, the approach in \cite{CarrozzaTanasa} seems to rely heavily on the particular case of rank three. 

The purpose of the present paper is to define and study a new consistent large $N$ scaling, working at all ranks and for all interactions, which enhances the BGR scaling for all non-melonic interactions. For rank three quartic interactions, this new scaling coincides with the one used by Carrozza and Tanasa in \cite{CarrozzaTanasa}. We shall consider $\smash{\text{O}(N)^{R}}$ invariant models made of real tensors of rank $R$; the complex tensors with $\smash{\text{U}(N)^{R}}$ symmetry, or mixed instances with, for example, $\smash{\text{O}(N)^{R_{1}}\times\text{U}(N)^{R_{2}}}$ symmetry, can be treated as special cases. The new scaling yields a new large $N$ limit, with expansion parameter $\smash{1/N^{\frac{1}{R-1}}}$. The expansion is no longer organized according to the degree of the Feynman graphs, but according to a new quantity that we call the \emph{index} of the graphs. This index has a very natural interpretation in terms of all the possible matrix models one can embed in the tensor model. The leading graphs, called \emph{generalized melons}, have index zero and form a larger class than the standard melons, which have degree zero. Their general classification remains a difficult open problem. One of our main results will be to provide such a classification in the particular case of the complete interaction vertex of order $R+1$, when $R$ is a prime number (Theorem \ref{theorem1}). We prove that the generalized melons in that case coincide with the \emph{mirror melons} defined in Sec.\ \ref{MirrorSec}. Besides, our construction singles out a new interesting family of models based on \emph{maximally single-trace} (MST) interactions, the complete interaction for $R$ prime being an example. The MST interactions generalize to tensor models the single-trace interactions of matrix models. Our scaling can be shown to be optimal for all MST terms.

\subsection{Matrix-tensor models and applications} 

Let us define more precisely the class of models on which we shall focus. It was of course noticed that to the independent indices of tensor models can correspond spaces of different dimensions, so that tensor models in fact generalize the Wishart theory of random \emph{rectangular} matrices. An interesting example, initiated in \cite{Bonzom1}, consists in singling out two indices out of $R=r+2$, $(a_{1}\cdots a_{R})=(ab\mu_{1}\cdots\mu_{r})$, and rewrite the tensor $T$ in terms of a matrix $X$, with $r$ additional tensor indices,
\be\label{tensortomatrix} T^{a_{1}\cdots a_{R}}=(X_{\mu_{1}\cdots\mu_{r}})^{a}_{\ b}\, .\ee
The natural symmetry is then $\text{O}(N)^{2}\times\text{O}(D)^{r}$, or other similar groups. We propose to call this class of models \emph{matrix-tensor models}; hence our title. Of course, we can always consider the case $D=N$, for which the theory reduces to that of a "hypercubic" rank $R=r+2$ tensor. However some non-trivial aspects of random matrix-tensors require a hierarchy between $N$ and $D$. As explained below, we shall typically assume $N >> D$ so that the limit $N \to \infty$ and $D \to \infty$ are performed in this order only.

The most general matrix-tensor models may include several matrix-tensors $X$, $Y$, etc., but this does not change our subsequent discussion, hence we can limit ourselves to a single $X$. We use real matrices with symmetry group $\text{O}(N)^{2}\times\text{O}(D)^{r}$, because this is the most general case, and we do not necessarily assume Hermiticity so that $X$ is not necessarily GOE but Wishart in the terminology of random matrices. The use of other symmetry groups, like unitary groups, simply amounts to consider a complex matrix $X$ and its conjugate $X^\dagger$. The corresponding study will be typically easier since the Feynman graphs in the complex case are bipartite and therefore consist in a strict subset of the ones in the the real case. The statistics of the matrix-tensors is irrelevant for the results presented in the paper, which apply equally to models containing bosonic and/or fermionic matrix-tensors. Some models containing bosons may be unstable, but, as is well-known, the large $N$ limit is still well-defined in this case because an exponentially bounded family of graphs survives \cite{Rivasseau:2016rgt}. Note that it is also easy to build stable models with bosons, including purely bosonic and supersymmetric examples \cite{ferra1,ferra2}. 

Since only the index structure matters when discussing the large $N$ and large $D$ limits, we can work in zero dimension most of the time without loss of generality. The general action we consider is then of the form
\be\label{genaction} S = ND^{r}\Bigl( \tr X_{\mu_{1}\cdots\mu_{r}}X^{T}_{\mu_{1}\cdots\mu_{r}} + \sum_{a}N^{1-t(\gB_{a})}\tau_{a}I_{\gB_{a}}(X)\Bigr)\ee
where the $\tau_{a}$ are coupling constants, the $I_{\gB_{a}}(X)$ are interaction terms and $t(\gB_{a})$ is the number of matrix traces in the interaction $I_{\gB_{a}}(X)$. The use of $\gB_{a}$ as labels is due to the fact that each $O(N)^2 \times O(D)^r$ invariant interaction term is associated to an $(r+2)$-regular edge-colored graph, also called \emph{bubble}, see Sec.\ \ref{LargeNDSec}. The explicit factors of $N$ have been chosen in such a way that the usual large $N$ scaling \`a la 't~Hooft corresponds to $N\rightarrow\infty$ at $\tau_{a}$ fixed. The global factor of $D^{r}$ in front of the action is the one inspired by the general scaling of tensors of rank $r$,
so that when $N=D$ we recover the $N^{(r+2)-1}$ scaling of tensors of rank $r+2$. The couplings $\tau_{a}$ will themselves have a non-trivial large $D$ scaling, discussed at length in Sec.\ \ref{LargeNDSec}.

A particularly interesting application of matrix-tensor models is as follows. By taking the large $N$ limit at fixed $D$ and $\tau_{a}$, one obtains the usual sum over planar diagrams of the matrix model \cite{tHooftplanar}. It is unnecessary to emphasize the prominent role played by the planar diagrams in modern theoretical physics, from QCD to black holes and string theory. In our framework, the additional parameter $D$ allows one to study the large $D$ expansion of the planar diagrams. As highlighted in \cite{ferra1,ferra2}, this large $D$ expansion is of fundamentally different nature depending on whether one uses the BGR scaling, as in \cite{Bonzom1}, or the new enhanced scaling we propose, for the couplings $\tau_{a}$. With the BGR scaling, it is straightforward to show that the large $N$ and the large $D$ limits commute, whereas with the new scaling, it is essential to take $N\rightarrow\infty$ first and $D\rightarrow\infty$ second for the limit to make sense. This fact turns out to have deep physical consequences. 

A simple intuitive explanation of why the difference between the two scalings is so crucial for physics can be given in the case of matrix-vector models, i.e.\ $r=1$ \cite{ferra1,ferra2}. If the large $N$ and large $D$ limits commute, one can actually take the limit $D\rightarrow\infty$ at fixed $N$ first. This is very similar to the standard large $D$ limit of vector models, reviewed for instance in \cite{vectorrev}. This limit selects a very restrictive class of Feynman graphs, the ``trees of bubbles,'' and the resulting physics is the same as in standard vector models. This is true already at fixed $N$ and remains of course true when $N$ goes to infinity, which eliminates even more diagrams. But vector models are much simpler than matrix models. The large $D$ approximation of the planar diagrams obtained in this way is thus bound to be a very poor approximation and it does not reproduce the most crucial physical properties of the full sum over planar diagrams. However, \emph{the situation is very different in our new scaling, for which large $N$ and large $D$ do not commute} \cite{ferra1}. Our new large $D$ expansion of matrix-vector models is totally different from the large $D$ expansion of vector models, because it includes a much wider class of Feynman diagrams. The truly remarkable point, emphasized in \cite{ferra1}, is that the main physical properties expected for the full sum over planar diagrams seem to be captured already at leading order. This property highlights the importance of the new enhanced scaling we study in the present paper. It provides another perspective on the deep relationship between matrix and tensor models and a new and reliable way to study physically relevant planar matrix models which were thought to be intractable before.

The research presented here connects with the ongoing effort to understand quantum models of black holes, following ideas first put forward by Kitaev \cite{Kitaev, poll,maldastan}. The so-called SYK model studied by Kitaev is non-standard because it uses quenched disorder, but it was pointed out by Witten in \cite{witten} that an ordinary quantum mechanics based on a colored tensor model shares the same basic properties. It was then realized in \cite{ferra1} that the basic structure of the Feynman graphs responsible for the remarkable properties of the SYK model was also relevant to planar matrix quantum mechanics, through the new large $D$ limit mentioned above. This made the link with holography and string theory clearer, since planar matrix models are singled out in this framework, the two indices of the matrices being the Chan-Paton factors associated with the two end points of open strings. The presence of an additional $\text{O}(D)$ index on the matrices was also naturally interpreted in \cite{ferra1}, as corresponding to the rotation group transverse to D-branes. The limit $D\rightarrow\infty$ is then physically similar to the large dimension limit of gravity studied in \cite{Emparan}. Our results provide a general framework to build a large class of solvable models with relevant properties to describe quantum black holes which, undoubtedly, have many interesting properties yet to be discovered.

\subsection{Plan of the paper}

In Sec.\ \ref{ColoredgraphSec} we introduce the mathematical tools to study the colored Feynman graphs associated to the matrix-tensor models. In particular, we define the important notion of index of a bubble with respect to a color. In Sec.\ \ref{LargeNDSec}, we define the large $N$ and large $D$ limits of matrix-tensor models by specifying a new scaling for the coupling constants. This new scaling enhances all the non-melonic interactions compared to the BGR scaling \cite{BGR}. Then, we show that the large $N$ and large $D$ limits are well-defined and that the large $D$ expansion is governed by the index of the Feynman graphs. We also show that our new scaling is optimal for all MST interactions. In Sec.\ \ref{applicationsSec}, we study in full detail the case of the complete interaction bubble for tensors of rank $R=r+2$. In particular, we prove a classification theorem for the generalized melons when $R$ is prime. We also briefly discuss $\text{O}(N)^{2}\times \text{O}(D)^{r}$ and $\text{U}(N)^{2}\times \text{O}(D)^{r}$ invariant matrix quantum mechanics based on Majorana and Dirac fermions respectively. Finally, we briefly point out a few interesting open problems in Sec.\ \ref{concSec}.

\section{\label{ColoredgraphSec}On colored graphs}

In this section we introduce the basic graph-theoretic tools that we shall need later. Most of the previous tensor model literature is restricted to the complex/bipartite case and $\uN$ symmetry groups. To treat the more general case of real matrix-tensors, we need to redefine many notions in the generalized framework of non-bipartite graphs. For example, our jackets can be non-orientable. We also introduce a central new object, \emph{the index with respect to a color}, and we derive its basic properties.

\subsection{\label{technical}Basic results on graphs}
\subsubsection{\label{defSec}Elementary definitions}

A graph is always denoted in curly letters like $\gG$ or $\gB$. The set of edges and vertices are denoted by $\mathscr E(\gG)$ and $\mathscr V(\gG)$, with cardinals $E(\gG)$ and $V(\gG)$ respectively. A \emph{bipartite} graph has a partition of the set of vertices $\mathscr V(\gG)=\mathscr V_{+}(\gG)\cup\mathscr V_{-}(\gG)$ such that edges join vertices in $\mathscr V_{+}(\gG)$ to vertices in $\mathscr V_{-}(\gG)$ only. A \emph{$d$-coloring} of $\gG$ is a surjective map $\mathscr E(\gG)\rightarrow \mathscr C$ where the set of colors $\mathscr C$ is isomorphic to $\{1,\ldots,d\}$. Unless explicitly stated otherwise, we assume that $\mathscr C =  \{1,\ldots,d\}$, in which case the colors are typically denoted by greek letters $\alpha$, $\beta$, etc., or that $\mathscr C =  \{0,\ldots,d-1\}$, in which case the color 0 is singled out and the colors $1,2,\ldots,d-1$ are denoted by latin indices $i$, $j$, etc. The number of edges of color $\alpha$ is $E_{\alpha}(\gG)$. The graph $\gG^{(\alpha_{1}\cdots \alpha_{p})}$ is obtained from $\gG$ by deleting all the edges of colors $\alpha_{1},\ldots ,\alpha_{p}$, whereas the graph $\gG_{(\alpha_{1}\cdots \alpha_{p})}$ is obtained from $\gG$ by keeping the edges of colors $\alpha_{1},\ldots ,\alpha_{p}$ and deleting all the others. The number of connected components of a graph $\gG$ is denoted by $G$ or more explicitly by $c(\gG)$. Similarly, we write $c(\gG^{(\alpha_{1}\cdots \alpha_{p})}) = G^{(\alpha_{1}\cdots \alpha_{p})}$ and  $c(\gG_{(\alpha_{1}\cdots \alpha_{p})}) = G_{(\alpha_{1}\cdots \alpha_{p})}$. The number of loops (independent cycles) of a graph $\gG$ is $L(\gG)=E(\gG)-V(\gG)+G$.

\subsubsection{\label{ineqSec}Connectivity inequalities and identities}

The following inequality will be very useful.
\begin{lemma}\label{ineqlem}
\be\label{ineq} G^{(\alpha\beta)}-G^{(\alpha)}-G^{(\beta)}+ G \geq 0\, .\ee
\end{lemma}
\noindent A straightforward generalization is obtained by replacing the colors $\alpha$ and $\beta$ by many colors $\alpha_{1},\ldots,\alpha_{p}$ and $\beta_{1},\ldots,\beta_{q}$ and by substituting $G^{(\gamma_{1}\cdots \gamma_{r})}$ to $G$ in \eqref{ineq},
\be\label{ineq2} G^{(\alpha_{1}\cdots\alpha_{p}\beta_{1}\cdots\beta_{q} \gamma_{1}\cdots \gamma_{r})}-G^{(\alpha_{1}\cdots\alpha_{p}\gamma_{1}\cdots \gamma_{r})}-G^{(\beta_{1}\cdots\beta_{q}\gamma_{1}\cdots \gamma_{r})}+ G^{(\gamma_{1}\cdots \gamma_{r})} \geq 0\, .\ee
These inequalities are valid in full generality, without putting any constraint on the coloring of the graph. 

\proof When one removes the lines of color $\alpha$ from $\gG$ and $\gG^{(\beta)}$, one creates $G^{(\alpha)}-G$ and $G^{(\alpha\beta)}-G^{(\beta)}$ new connected components, respectively. But a graph that splits when the lines of colors $\beta$ are not taken into account may remain connected otherwise. This implies that $G^{(\alpha)}-G\leq G^{(\alpha\beta)}-G^{(\beta)}$, which is \eqref{ineq}. Another  argument amounts to noting that the left-hand side of \eqref{ineq} matches with the number of loops in the abstract bipartite graph $\gG_{\alpha,\beta}$ built as follows: the $+$ and $-$ vertices of $\gG_{\alpha,\beta}$ are the connected components of $\gG^{(\alpha)}$ and $\gG^{(\beta)}$ respectively and an edge joins a $+$ to a $-$ vertex for each connected component of $\smash{\gG^{(\alpha\beta)}}$ included in both vertices. It is straightforward to check that $\gG_{\alpha,\beta}$ has $G$ connected components. Moreover, by construction, $V(\gG_{\alpha,\beta})=G^{(\alpha)}+G^{(\beta)}$ and $E(\gG_{\alpha,\beta})=G^{(\alpha\beta)}$; thus $L(\gG_{\alpha,\beta})=G^{(\alpha\beta)}-G^{(\alpha)}-G^{(\beta)}+G\geq0$.\qed

For later purposes, it will be convenient to use the above inequalities in the following form. We single out the color 0 and label the other colors with latin indices. The quantity
\be\label{Deltadef} \Delta_{0} G^{(i_{1}\cdots i_{p})} = G^{(0i_{1}\cdots i_{p})}-G^{(i_{1}\cdots i_{p})}\ee
represents the number of new connected components that are created when one removes the lines of color 0 from $\gG^{(i_{1}\cdots i_{p})}$. One can decompose
\be\label{Deltadec}\begin{split}
\Delta_{0} G^{(i_{1}\cdots i_{p})} - \Delta_{0} G & = G^{(0i_{1}\cdots i_{p})}-G^{(i_{1}\cdots i_{p})} - G^{(0)}+ G\\
& = \bigl(\Delta_{0} G^{(i_{1}\cdots i_{p})}-\Delta_{0} G^{(i_{1}\cdots i_{p-1})}\bigr) \\
& \hskip .4cm  + \bigl(\Delta_{0} G^{(i_{1}\cdots i_{p-1})}-\Delta_{0} G^{(i_{1}\cdots i_{p-2})}\bigr)
 + \cdots + \bigl(\Delta_{0} G^{(i_{1})}-\Delta_{0} G\bigr)
\end{split}\ee
as a sum of terms that are all positive according to \eqref{ineq2}. In particular, the condition 
\be\label{vcond1} \Delta_{0} G^{(i_{1}\cdots i_{p})} = \Delta_{0} G\ee
is equivalent to the conditions 
\be\label{vcond2}\Delta_{0} G^{(i_{1}\cdots i_{k})}=\Delta_{0} G^{(i_{1}\cdots i_{k-1})}\ee
for all $1\leq k\leq p$. It will be useful to sum the decomposition \eqref{Deltadec} over all the possible indices $i_{1},\ldots,i_{p}$. If we introduce the positive integers
\be\label{ncon1} \delta_{0;p}(\gG)  =\sum_{i_{1}<i_{2}< \cdots< i_{p}} \bigl(\Delta_{0} G^{(i_{1}\cdots i_{p})} - \Delta_{0} G\bigr)=
\sum_{i_{1}<i_{2}< \cdots< i_{p}}\bigl( G^{(0 i_{1}\cdots i_{p})}-G^{(i_{1}\cdots i_{p})}-G^{(0)}+G\bigr)\ee
and
\begin{multline}\label{nconnex1} \tilde\delta_{0;p} (\gG)  = \sum_{i_{1}< i_{2}< \cdots < i_{p}}
\bigl(\Delta_{0} G^{(i_{1}\cdots i_{p})}-\Delta_{0} G^{(i_{1}\cdots i_{p-1})}\bigr)\\=
\sum_{i_{1}< i_{2}< \cdots < i_{p}}\bigl( G^{(0 i_{1}\cdots i_{p})}-G^{(i_{1}\cdots i_{p})}-G^{(0 i_{1}\cdots i_{p-1})}+G^{(i_{1}\cdots i_{p-1})}\bigr) \, ,
\end{multline}
then \eqref{Deltadec} yields
\be\label{conid} \delta_{0;p}(\gG)  = \frac{1}{p!(d-p-1)!}\sum_{k=1}^{p}k!(d-k-1)!\, \tilde\delta_{0;k}(\gG)\, .\ee
The condition $\delta_{0;p}=0$ is equivalent to \eqref{vcond1} for all possible indices $i_{1},\ldots,i_{p}$, which is also equivalent to $\tilde\delta_{0;k}=0$ and to \eqref{vcond2} for all possible indices $i_{1},\ldots,i_{k}$ and $1\leq k\leq p$.

Finally, let us note that the formula \eqref{ncon1} for $\delta_{0;p}$ can also be written as
\be\label{ncon2} \delta_{0;d-1-p}(\gG) =
\sum_{i_{1}<i_{2}< \cdots< i_{p}}\bigl( G_{(i_{1}\cdots i_{p})}-G_{(0i_{1}\cdots i_{p})}-G^{(0)}+G\bigr)\, .\ee
For example, we obtain
\begin{align}\label{del1}\delta_{0;d-1}(\gG) & = V(\gG) - E_{0}(\gG) - G^{(0)} + G\, ,\\
\label{del2}
\delta_{0;d-2}(\gG) & =\sum_{i}\bigl( E_{i}(\gG) - G_{(0i)} - G^{(0)} + G\bigr)\, ,\\
\label{del3}
\delta_{0;d-3}(\gG) & =\sum_{i<j}\bigl(G_{(ij)} - G_{(0ij)}-G^{(0)}+G\bigr)\, .
\end{align}

\subsection{\label{bubjackSec}Bubbles, jackets and degree}

\subsubsection{Bubbles}

A \emph{$d$-bubble} $\gB$ is a $d$-colored regular graph, i.e.\ all the vertices have the same valency $d$ and the $d$ edges incident to a given vertex carry the $d$ possible distinct colors. In particular,
\be\label{EVrel} 2 E(\gB) = d\, V(\gB) = 2 d\, E_{\alpha}(\gB)\, .\ee
Note that, if $\gB$ is a $d$-bubble, $\gB^{(\alpha_{1}\cdots \alpha_{p})}$ and $\gB_{(\alpha_{1}\cdots \alpha_{p})}$ are $(d-p)$ and $p$-bubbles, respectively, for any $p\leq d$. Given two distinct colors $\alpha$ and $\beta$, a \emph{face} of colors $\alpha$ and $\beta$, also called an $(\alpha\beta)$-face, is defined to be a cycle of $\gB$ made of edges of alternating colors $\alpha$ and $\beta$. Equivalently, the $(\alpha\beta)$-faces are the connected components of $\gB_{(\alpha\beta)}$. The total number of $(\alpha\beta)$-faces is $F_{\alpha\beta}(\gB)=B_{(\alpha\beta)}$. The total number of faces is $F(\gB) = \sum_{\alpha<\beta}F_{\alpha\beta}(\gB)$.

\subsubsection{Jackets}

In general, there are many ways to draw an abstract graph on a surface. To each such drawing, or \emph{embedding}, is associated a standard ribbon graph \`a la 't~Hooft. The distinct embeddings are labeled by a choice of cyclic ordering of the incident edges at each vertex of the graph together with a choice of ``signature'' for each edge. In the ribbon graph representation, edges of signatures $+$ and $-$ are represented by untwisted and twisted ribbons respectively. If the embedded graph has at least one cycle with an odd number of negative signature edges, the surface is non-orientable. If one reverses the cyclic ordering at a given vertex and at the same time flips the signatures of all the incident edges at this vertex, an operation that we shall call a ``local switch,'' one obtains an equivalent embedding. To any given embedding of the graph is associated the genus of the corresponding surface. Note that, when the graph has several connected components, we define the genus to be the sum of the genera of its connected components.

There is a natural set of embeddings associated to any $d$-bubble. We pick a cycle $\sigma\in S_{d}$ and a partition $\pi$ of the set of vertices of the graph into two disjoint subsets $\mathscr V_{+}(\gB)$ and $\mathscr V_{-}(\gB)$. The vertices in $\mathscr V_{+}(\gB)$ and $\mathscr V_{-}(\gB)$ are called filled and unfilled, respectively. The embedding associated with the pair $(\sigma,\pi)$ is then defined as follows. The colored edges are cyclically ordered clockwise according to $\sigma$ around each unfilled vertex and anticlockwise around each filled vertex. The signature of an edge joining two vertices of the same type (filled or unfilled) is chosen to be $-$ and the signature of an edge joining two vertices of different types is chosen to be $+$. The resulting ribbon graph is called a \emph{jacket} associated with $\gB$ and is denoted as $\gJ(\gB;\sigma,\pi)$. To any jacket of $\gB$ is associated a genus $g(\gJ(\gB;\sigma,\pi))= g(\gB;\sigma,\pi)$. Jackets have the following properties.

\begin{proposition}\label{jacketprop}
i) A jacket $\gJ(\gB;\sigma,\pi)$ does not depend on the choice of the partition $\pi$. ii) $\gJ(\gB,\sigma)=\gJ(\gB,\sigma^{-1})$. We can thus associate $\smash{\frac{1}{2}(d-1)!}$ distinct jackets to a given $d$-bubble. iii) If one jacket $\gJ(\gB,\sigma)$ is orientable, then all the jackets $\gJ(\gB,\sigma')$, for any $\sigma'$, are orientable. We then say that the bubble $\gB$ itself is \emph{orientable}. iv) A bubble is orientable if and only if the underlying graph is bipartite. In particular, planar jackets $\gJ(\gB,\sigma)$ can exist only if $\gB$ is bipartite.
\end{proposition} 
\proof i) If one changes the type of a given vertex, then, by definition, the new jacket is obtained from the old one by a local switch. We can thus denote the jackets simply by $\gJ(\gB,\sigma)$, even though, in practice, when one draws the graph, one chooses a partition $\pi$. ii) This is a consequence of i), because changing $\sigma$ to $\sigma^{-1}$ is equivalent to flipping the type of all the vertices of the graph. iii) If $\gJ(\gB,\sigma)$ is orientable, then it is well-known that, modulo local switches, it has a representation with untwisted ribbons only. The underlying graph is thus manifestly bipartite. We can then use the partition $\pi$ associated with the bipartite structure to find a manifestly orientable embedding for any jacket $\gJ(\gB;\sigma')$. iv) Immediately follows from the proof of iii).\qed

\subsubsection{Degree}

The degree of a $d$-bubble $\gB$, for $d\geq 3$, is defined as the sum of the genera of its jackets \cite{colored} 
\be\label{degreedef} \deg\gB = \frac{1}{2}\sum_{\text{cycles}\ \sigma\in S_{d}} g(\gB;\sigma)\, .\ee
The factor of $\frac{1}{2}$ simply takes into account that $\gJ(\gB,\sigma)=\gJ(\gB,\sigma^{-1})$. By construction, the degree is a non-negative half-integer; a stronger result can actually be proven, see \eqref{degreeN}. Note that the degree of a multiply-connected bubble is the sum of the degrees of the connected components. The case of 3-bubbles is particularly simple: there is only one jacket and thus the degree coincides with the genus of the associated ribbon graph.

At this stage, one has in principle two distinct notions of faces. On the one hand, we have the $(\alpha\beta)$-faces of the bubble $\gB$ as defined previously. On the other hand, we have the usual faces associated with the ribbon graphs $\gJ(\gB,\sigma)$. There is a fundamental relation between the two notions given by the following proposition.
\begin{proposition}\label{faceprop} The faces of the ribbon graph $\gJ(\gB,\sigma)$ are in one-to-one correspondence with the subset of $(\alpha\beta)$-faces of $\gB$ satisfying $\beta=\sigma(\alpha)$ or $\alpha=\sigma(\beta)$.
\end{proposition}
\proof This is a direct consequence of the precise definition of the jackets. In particular, it relies on the rule for twisting or untwisting the ribbons.\qed

One can then derive a very useful generalization of the usual Euler's formula relating the genus to the number of faces, vertices and edges of a ribbon graph.
\begin{proposition}\label{degreeformprop} The degree satisfies
\be\label{degreeformula} (d-1)c(\gB) - \frac{2}{(d-2)!}\deg\gB = F(\gB) - \frac{1}{4}(d-1)(d-2) V(\gB)\ee
and in particular
\be\label{degreeN} \frac{2}{(d-2)!}\deg\gB\in\mathbb N\, .\ee
\end{proposition}
\proof Eq.\ \eqref{degreeformula} follows from a simple counting of the faces using Proposition \ref{faceprop}. Taking into account the fact that $V(\gB)$ is always even, see \eqref{EVrel}, Eq.\ \eqref{degreeformula} implies \eqref{degreeN}.\qed

Finally, we shall need the following crucial lemma, which is a direct generalization of Lemma 7 in \cite{Gurau:2011tj}.
\begin{lemma}\label{degdeglem} For any $d\geq 4$ and any choice of color $\alpha$,
\be\label{degineq} \deg\gB \geq (d-1)\deg\gB^{(\alpha)}\, .\ee
\end{lemma}
\proof To any jacket $\gJ(\gB,\sigma)$ is associated the jacket $\gJ(\gB^{(\alpha)},\sigma^{(\alpha)})$, where $\sigma^{(\alpha)}$ is the cycle obtained from $\sigma$ by deleting the color $\alpha$. The jacket $\gJ(\gB^{(\alpha)},\sigma^{(\alpha)})$ is obtained from $\gJ(\gB,\sigma)$ by deleting the ribbons associated with the edges of color $\alpha$ in $\gB$. It is well-known that this operation cannot increase the genus of the ribbon graph, $g(\gB,\sigma)\geq g(\gB^{(\alpha)},\sigma^{(\alpha)})$. Summing this inequality over all cycles $\sigma$ yields \eqref{degineq}.\qed

\subsection{\label{indsec}The index of a bubble with respect to a color}

\subsubsection{Definition}

In this subsection, we use again the set of colors $\mathscr C=\{0,1,\ldots,d-1\}$, singling out the color 0 for convenience and using latin indices $i$, $j$, etc., to label the colors from $1$ to $d-1$, but not 0.

The \emph{index} of a $d$-bubble $\gB$, $d\geq 4$, with respect to the color $0$ is defined to be
\be\label{inddef} \ind_{0}\gB = \frac{1}{(d-3)!}\Bigl(\deg\gB - (d-1)\deg \gB^{(0)}\Bigr)\, .\ee
From \eqref{degreeN} and \eqref{degineq}, we deduce that the index is a non-negative half-integer,
\be\label{ind0N} \ind_{0}\gB\in\frac{1}{2}\,\mathbb N\, .\ee
If $\gB$ is bipartite, the decomposition formula \eqref{fundid0} below actually shows that $\ind_{0}\gB$ is an integer. If the graph is multiply-connected, its index is the sum of the indices of the connected components.

\subsubsection{The decomposition formula}

\begin{proposition} \label{fundidth} (First form of the decomposition formula) The index can be expressed as a sum of manifestly positive contributions as
\be\label{fundid0} \ind_{0}\gB = \sum_{i<j}\bigl(g(\gB_{(0ij)}) +  F_{ij}(\gB) -B_{(0ij)}-B^{(0)}+B\bigr)\, .\ee
\end{proposition}
Equivalently, using \eqref{del3}, Eq.\ \eqref{fundid0} can be rewritten as
\be\label{fundid} \ind_{0}\gB = \sum_{i<j}g(\gB_{(0ij)}) + \delta_{0;d-3}(\gB)\, .\ee
One can also use \eqref{conid} to obtain an alternative expression. For example, for $d=4$, we get
\be\label{fid1} \ind_{0}\gB = \sum_{i=1}^{3}g(\gB^{(i)}) + \sum_{i=1}^{3}\bigl(
B^{(0i)}-B^{(i)}-B^{(0)}+B\bigr) \, , \ee
and for $d=5$,
\begin{multline}\label{fid2} \ind_{0}\gB = \sum_{1\leq i<j\leq 4}g(\gB^{(ij)}) + \frac{3}{2}\sum_{i=1}^{4}\bigl(
B^{(0i)}-B^{(i)}-B^{(0)}+B\bigr) \\+ \sum_{1\leq i < j\leq 4}\bigl(
B^{(0ij)}-B^{(ij)}-B^{(0i)}+B^{(i)}\bigr)\, .
\end{multline}

The decompostion formula \eqref{fundid0} is central in our work. It has a very natural physical interpretation, which will be explained in Sec.\ \ref{LargeNDSec}, Eq.\ \eqref{fundid1}. Actually, an equally valid presentation would consist in using \eqref{fundid0} to define the index and then derive \eqref{inddef}, instead of the other way around.

\proof The proof uses the following lemma, which is the analogue of \eqref{degreeformula} for the index.
\begin{lemma}\label{indexlem} The index can be expressed as
\begin{multline}\label{indexplicit} \ind_{0}\gB = \frac{1}{2}(d-1)(d-2) \bigl(B-B^{(0)}\bigr)+\frac{1}{8}(d-1)(d-2) V(\gB)\\
+ \frac{1}{2}\sum_{i<j}F_{ij}(\gB)-\frac{1}{2}(d-2)\sum_{i} F_{0i}(\gB)\, .
\end{multline}
\end{lemma}
The expression \eqref{indexplicit} is obtained by using \eqref{degreeformula} for the $d$-bubble $\gB$ and the $(d-1)$-bubble $\gB^{(0)}$ and decomposing
\be\label{facedec} F(\gB) = \sum_{i=1}^{d-1}F_{0i}(\gB) + \sum_{i<j}F_{ij}(\gB)\, ,\quad F(\gB^{(0)}) = \sum_{i<j}F_{ij}(\gB)\, .\ee
To proceed further, we use Euler's formula for all the 3-bubbles $\gB_{(0ij)}$. Summing the resulting $\frac{1}{2}(d-1)(d-2)$ equations yields
\be\label{indstep1} \sum_{i<j} \bigl(2 B_{(0ij)} - 2 g(\gB_{(0ij)})\bigr) =(d-2)\sum_{i}F_{0i} (\gB) + \sum_{i<j}F_{ij}(\gB) -\frac{1}{4}(d-1)(d-2) V(\gB)\, .\ee
We then use this result to eliminate $V(\gB)$ from \eqref{indexplicit} to get \eqref{fundid0}.\qed

\subsubsection{Addition formulas}

\begin{figure}[h]
\centerline{\includegraphics[width=6in]{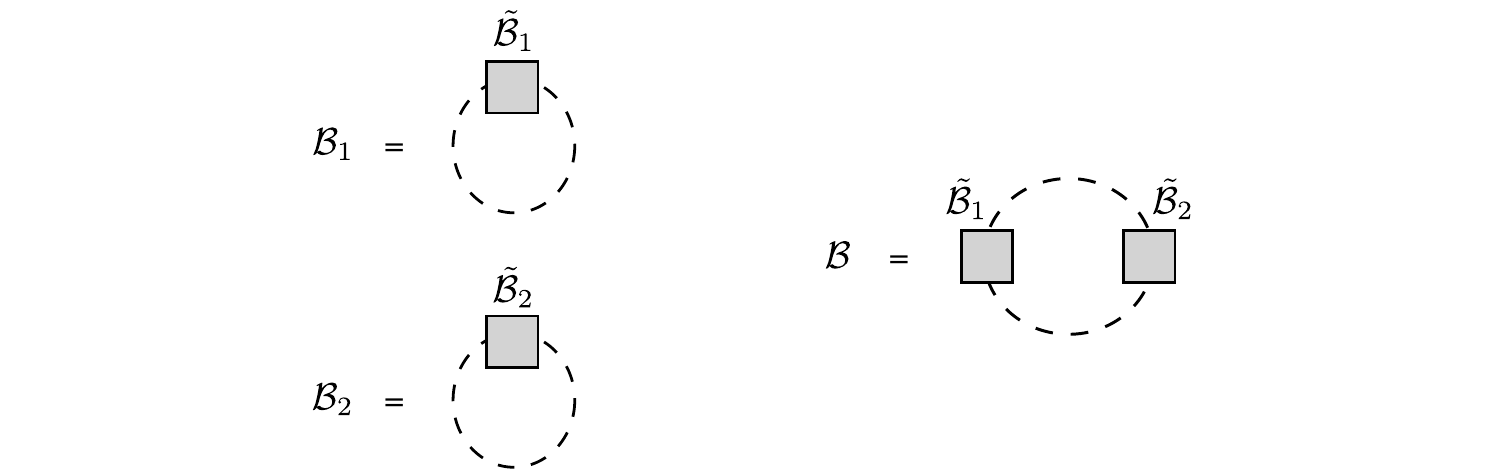}}
\caption{Construction of a new bubble $\gB$ from bubbles $\gB_{1}$ and $\gB_{2}$, such that $\deg \gB = \deg \gB_{1}+\deg\gB_{2}$ and $\ind_{0}\gB  = \ind_{0}\gB_{1}+\ind_{0}\gB_{2}$ (Proposition \ref{addprop}). Dashed lines represent edges of color 0.}\label{additionfig}
\end{figure}

\begin{proposition}\label{addprop} (Addition formulas) Consider two $d$-bubbles $\gB_{1}$ and $\gB_{2}$ and the so-called two-point graphs $\tilde\gB_{1}$ and $\tilde\gB_{2}$ obtained by cutting open any edge of color 0 in $\gB_{1}$ and $\gB_{2}$ respectively. Build a new $d$-bubble $\gB$ by gluing the open edges of color 0 in $\tilde\gB_{1}$ and $\tilde\gB_{2}$ as depicted in Fig.\ \ref{additionfig} (note that there are in general two inequivalent ways to perform this gluing). Then
\begin{align}\label{adddeg} \deg \gB &= \deg \gB_{1}+\deg\gB_{2}\\\label{addind}
\ind_{0}\gB & = \ind_{0}\gB_{1}+\ind_{0}\gB_{2}\, .
\end{align}
\end{proposition} 
\proof By construction, $c(\gB)=c(\gB_{1})+c(\gB_{2})-1$, $V(\gB)=V(\gB_{1})+V(\gB_{2})$ and $F_{ij}(\gB)=F_{ij}(\gB_{1})+F_{ij}(\gB_{2})$. Moreover, the two $(0i)$-faces, for some color $i$, in $\gB_{1}$ and $\gB_{2}$ going through the edges of color 0 that are cut open are joined in a unique $(0i)$-face in $\gB$, whereas the other $(0i)$-faces remain unchanged. This yields $F_{0i}(\gB)=F_{0i}(\gB_{1})+F_{0i}(\gB_{2})-1$. Overall, we thus get $F(\gB)=F(\gB_{1})+F(\gB_{2})-(d-1)$. Eq.\ \eqref{adddeg} then follows using \eqref{degreeformula} for the bubbles $\gB$, $\gB_{1}$ and $\gB_{2}$. Eq.\ \eqref{addind} can be proven by a similar reasoning using \eqref{indexplicit} or from the definition \eqref{inddef} using \eqref{adddeg} and the trivial result $\deg\gB^{(0)}=\deg\gB_{1}^{(0)}+\deg\gB_{2}^{(0)}$.\qed

\begin{figure}
\centerline{\includegraphics[width=6in]{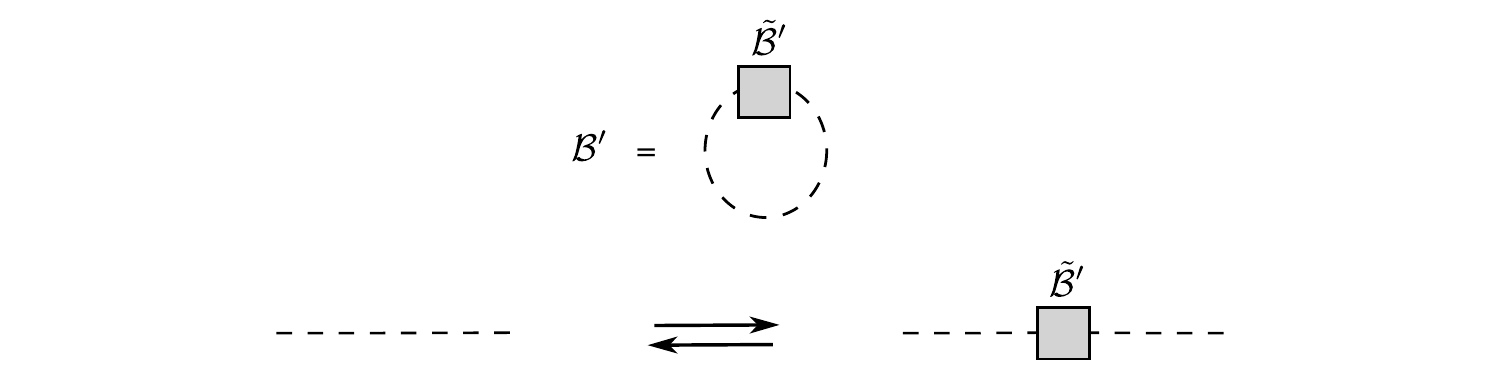}}
\caption{Bubble $\gB'$ insertion (from left to right) and contraction (from right to left) on an edge of color 0 in a bubble $\gB$. According to Proposition \ref{addprop}, the insertion (contraction) increases (decreases) the degree and the index of $\gB$ by $\deg\gB'$ and $\ind_{0}\gB'$ respectively.}\label{insertfig}
\end{figure}

The operation of \emph{bubble insertion} of a $d$-bubble $\gB'$ is defined as the replacement of an edge of color 0 in a given $d$-bubble $\gB$ by the two-point graph $\tilde\gB'$ obtained from $\gB'$, as depicted in Fig.\ \ref{insertfig}. Proposition \ref{addprop} implies that $\deg\gB\mapsto\deg\gB + \deg\gB'$  and $\ind_{0}\gB\mapsto\ind_{0}\gB + \ind_{0}\gB'$ under this operation. The inverse operation is called a \emph{bubble contraction}.

\subsubsection{\label{gmSec}Generalized melons and generalized melonic moves}

We call the graphs of index zero \emph{generalized melons}. Since the index of a graph is the sum of the indices of its connected components, a graph is of index zero if and only if all its connected components have index zero. For a connected graph $\gB$, the decomposition formula \eqref{fundid} implies that $\ind_{0}\gB=0$ is equivalent to the conditions
\begin{align}\label{ind01} & g(\gB_{(0ij)}) = 0\, ,\\ \label{ind02}
& F_{ij}(\gB) - B_{(0ij)} = B^{(0)}-1\, ,
\end{align}
for all choices of pairs of colors $(i,j)$. The first condition simply says that the uniquely defined ribbon graphs, obtained by keeping any combination of three colors $(0,i,j)$ and removing all the others from the original bubble, are planar. As for the second condition, it can also be written as
\be\label{zeroindex1} \Delta_{0}B^{(i_{1}\cdots i_{d-3})}=B^{(0i_{1}\cdots i_{d-3})}-B^{(i_{1}\cdots i_{d-3})}=\Delta_{0}B=B^{(0)}-1\, ,\ee
for all choices of indices $i_{1},\ldots,i_{d-3}$, or, as explained in \ref{ineqSec}, as
\be\label{zeroindex}\Delta_{0}B^{(i_{1}\cdots i_{p})}= B^{(0 i_{1}\cdots i_{p})}-B^{(i_{1}\cdots i_{p})}=\Delta_{0}B^{(i_{1}\cdots i_{p-1})}=B^{(0 i_{1}\cdots i_{p-1})}-B^{(i_{1}\cdots i_{p-1})}\, ,\ee
for all choices of indices and $1\leq p\leq d-3$. A convenient way to understand these conditions is to consider a graph for which the set of $B^{(0)}$ connected components of $\gB^{(0)}$ are given. The lines of color 0 must then be adjusted in such a way that $B_{(0ij)}$ takes its maximum possible value, that is to say, the graph $\gB_{(0ij)}$ is maximally disconnected, for all pairs $(i,j)$. Both conditions \eqref{ind01} and \eqref{ind02} are shown to have a very natural physical interpretation in the next section.

Note that the inequality \eqref{degineq} implies that ordinary melons, i.e.\ graphs of degree zero, are also generalized melons. But the converse is not true: the class of generalized melons is much larger. In particular, the connected components of $\gB^{(0)}$ can be non-melonic bubbles. 

Proposition \ref{addprop} implies that inserting or contracting a generalized melon on an edge of color 0 in an arbitrary bubble $\gB$ does not change the index of $\gB$. These operations will be called \emph{generalized melonic moves}. A useful consequence is that, from any given generalized melon, one can immediately construct an infinite family of generalized melons by repeated generalized melonic insertions. 

\section{\label{LargeNDSec}Large \texorpdfstring{$N$}{N} and large \texorpdfstring{$D$}{D}}
\subsection{\label{bubFeynmanSec}Bubbles and Feynman graphs}

The interaction terms of matrix-tensor models, entering the action \eqref{genaction}, are in one-to-one correspondence with $R$-bubbles $\gB_{a}$: each tensor field is associated to a vertex, each index of the tensors is associated with a color and the colored edges are drawn according to the way the indices are contracted in the interaction term. The fact that one gets a $R$-bubble through this procedure is equivalent to the $\text{O}(N)^{2}\times\text{O}(D)^{r}$ invariance. To each interaction term $\gB_{a}$, we associate:
\begin{itemize} 
\item the number of traces appearing in the interaction, which is $t(\gB_{a}) = F_{12}(\gB_{a})$,
\item the number of connected components of the bubble, which is $c(\gB_{a})=B_{a}$; note that $t(\gB_{a})\geq c(\gB_{a})$,
\item the degree $\deg \gB_{a}$ of the bubble.
\end{itemize} 
A typical special case to consider are models with only single-trace interactions, for which $t(\gB_{a})=1$ and thus $c(\gB_{a})=1$ as well. In the following, we denote the colors 1 to $R$ by lower case latin indices $i$, $j$, etc., and the $r$ colors $3$ to $R$ by upper case latin indices $I$, $J$, etc. 

Vacuum Feynman graphs are themselves in one-to-one correspondence with $(R+1)$-bubbles, the color 0 being associated with the propagators. We shall mostly restrict ourselves to vacuum graphs for simplicity, but associated Feynman graphs with external sources can be defined and studied straightforwardly in a standard way. 

\subsection{Large \texorpdfstring{$N$}{N} and large \texorpdfstring{$D$}{D} scalings}

To define the large $N$ and large $D$ limits of the models, we need to specify how the coupling constants $\tau_{a}$ in the action \eqref{genaction} scale with $N$ and $D$. The scaling must be such that the associated large $N$ and large $D$ limits exist and are non-trivial. Unlike the cases of ordinary vector and matrix models, there exist in general several natural choices of interesting scalings. 

\noindent\emph{The BGR scaling}: it is straightforward to generalize the scaling used in \cite{BGR} to the symmetry group $\text{O}(N)^{2}\times\text{O}(D)^{r}$ (instead of $\text{U}(N)^{R}$) and to the case of multiply-connected interaction bubbles $\gB_{a}$. One defines the couplings $\mu_{a}$ in terms of the couplings $\tau_{a}$ appearing in the action \eqref{genaction} by
\be\label{BGRcouplings} \tau_{a}=D^{t(\gB_{a})-c(\gB_{a})-\frac{2}{r!}\deg \gB_{a}}\mu_{a}\, .\ee
The limits $N\rightarrow\infty$ and $D\rightarrow\infty$ are then formulated by keeping the $\mu_{a}$ fixed. It is easy to show (see below) that the limits $N\rightarrow\infty$ and $D\rightarrow\infty$ commute in this case. When $D=N$, one obtains the usual large $N$ tensorial expansion organized according to the degree of the Feynman graphs \cite{BGR}. The leading graphs are the melonic, degree-zero graphs. 
They can be fully classified and enumerated \cite{melons}. 

However the family of melonic graphs is quite restricted from the point of view of matrix-tensor models. In particular, using \eqref{degineq}, it is clear that at leading order only the interaction vertices that are themselves melonic can contribute. As emphasized in \cite{ferra1}, the scarcity of the leading graphs implies that the scaling \eqref{BGRcouplings} yields a physics which has similarities with the large $D$ limit of vector models.
%
%
%
%

It seems unlikely that the melonic
interactions of tensor models can alone capture the SYK physics; for instance the quartic melonic interaction generates a tadpole at leading order in the two-point function, which is not bi-local in time like in SYK and thus behaves exactly like a vector model. As stressed in \cite{ferra1}, this is why the tetraedric interaction and the scaling of \cite{CarrozzaTanasa}, unlike uncolored melons with BGR scaling, play a crucial role in works like \cite{klebanov}. 

This motivates us to look for enhanced scalings which, like the degree, work for any interaction, but which generalize the tetraedric scaling of \cite{CarrozzaTanasa}. The goal is to obtain large $N$ and large $D$ limits that include many more Feynman graphs and are thus able to capture the essential aspects of the sum over planar diagrams, in particular the SYK/black hole physics. As we shall see, our new scaling is going to be optimal for a large class of interactions (but not all).

\noindent\emph{The new scaling}: we define a new large $N$ and large $D$ limit for which the couplings $\la_{a}$, defined by
\be\label{newcouplings} \boxed{\tau_{a}= D^{t(\gB_{a})-c(\gB_{a})+\frac{2}{(r+1)!}\deg \gB_{a}}\la_{a}}\, ,\ee
are held fixed. Note that
\be\label{newvsold} \mu_{a} = D^{\frac{2(r+2)}{(r+1)!}\deg\gB_{a}}\la_{a}\, .\ee
Taking $D\rightarrow\infty$ at fixed $\la_{a}$ amounts to an infinite amplification of all the non-melonic ($\deg\gB_{a}>0$) interactions with respect to the BGR scaling, as announced. In spite of this enhancement, we are going to demonstrate below that the large $N$ and large $D$ limits still exist, with a crucial new feature first seen in \cite{ferra1}: the two limits no longer commute. One should first take $N\rightarrow\infty$ and then $D\rightarrow\infty$ at each order in the $1/N$ expansion to recover the desired picture of a large $D$ expansion within each term of 't~Hooft's expansion. Moreover, the large $N$ and large $D$ expansions are no longer governed by the degree of the Feynman graphs with expansion parameter $1/D$, as in \cite{BGR}, but by their index, defined in Sec.\ \ref{indsec}, with expansion parameter $\smash{1/D^{\frac{1}{r+1}}}$.

\subsection{\label{largeNDSec}Large \texorpdfstring{$N$}{N} and large \texorpdfstring{$D$}{D} expansions}

Let us denote by $v$, $p$, $f$ and $\varphi$ the number of vertices, propagators, $\text{O}(N)$ index loops and $\text{O}(D)$ index loops in a given connected Feynman diagram. With the action \eqref{genaction} and scaling \eqref{newcouplings}, to each propagator is associated a factor $1/(ND^{r})$, to each vertex corresponding to the interaction term $\gB_{a}$ a factor $$N^{2-t(\gB_{a})}D^{r+t(\gB_{a})-c(\gB_{a})+\frac{2}{(r+1)!}\deg\gB_{a}}\, ,$$ to each $\text{O}(N)$ index loop a factor $N$ and to each $\text{O}(D)$ index loop a factor $D$. Overall, the amplitude of a Feynman diagram is thus proportional to
\be\label{Feyn1} N^{2-h}D^{r+h-\frac{\ell}{r+1}}=\frac{1}{(ND^{r})^{p}}N^{2v-\sum_{a}t(\gB_{a})+f}D^{rv+\sum_{a}\left(t(\gB_{a})-c(\gB_{a})+\frac{2}{(r+1)!}\deg\gB_{a}\right)+\varphi}\, .\ee
\subsubsection{Factor of \texorpdfstring{$N$}{N}} \label{factorN}

\noindent\emph{General case}

The computation of the factor of $N$ is a standard exercice for multi-trace matrix models, see e.g.\ \cite{ferra2}, but we repeat here the argument because the result is important for our purposes. From \eqref{Feyn1}, we read off
\be\label{hf1} h = 2+p-2v+\sum_{a}t(\gB_{a})-f\, .\ee
We then consider the matrix model ribbon graph obtained by removing all the lines of colors $3$ to $r+2$. Because some of the vertices in the connected Feynman graph can be multi-trace, this ribbon graph is not necessarily connected. Each multi-trace vertex effectively yields $t(\gB_{a})$ distinct single-trace vertices in the ribbon graph. The total number of effective single-trace vertices is thus $\tilde v = \sum_{a}t(\gB_{a})$. The number of connected components matches the number of connected components $B_{(012)}$ of the three-bubble $\gB_{(012)}$. Euler's formula then yields the genus $g$ of the ribbon graph,
\be\label{Eulerfat1} 2B_{(012)}-2g = f-p+\tilde v = f-p+\sum_{a}t(\gB_{a})\, .\ee
By using the following relations between the Feynman graph data and the colored graph data,
\be\label{fcrel1} \sum_{a}t(\gB_{a}) = F_{12}(\gB)\, ,\quad p = E_{0}(\gB) = \frac{1}{2}V(\gB)\, ,\quad f=F_{01}(\gB) + F_{02}(\gB)\, ,\ee
we also get, for the genus of the three-colored graph $\gB_{(012)}$, the relation
\be\label{B012genus}\begin{split}
2B_{(012)}-2g(\gB_{(012)}) & = F(\gB_{(012)})-E(\gB_{(012)}) + V(\gB_{(012)})\\ &
= F_{01}(\gB) + F_{02}(\gB)+F_{12}(\gB)-\frac{1}{2}V(\gB)\\
& = f-p+\sum_{a}t(\gB_{a})
\end{split}\ee
and thus, not surprisingly,
\be\label{ggrel} g = g(\gB_{(012)})\, .\ee
Plugging the above formulas in \eqref{hf1} yields
\be\label{hf2} \frac{h}{2} = g+1-B_{(012)}+\sum_{a}\bigl(t(\gB_{a})-1\bigr)\, .\ee

It is important to realize that the right-hand side of \eqref{hf2} is the sum of two positive terms,
\be\label{posterms1} g\geq 0\, ,\quad 1-B_{(012)}+\sum_{a}\bigl(t(\gB_{a})-1\bigr)\geq 0\, .\ee
The positivity of the second term in \eqref{posterms1} can be seen as a special case of the connectivity inequality \eqref{ineq}.\footnote{It is straightforward to check that the four-colored graph $\smash{\hat\gB_{(012)}}$, built from $\gB_{(012)}$ by adding lines of a new color $\hat 0$ joining together all the single-trace pieces within the multi-trace vertices, is such that $\hat B_{(012)}^{(\hat 0 0)}-\hat B_{(012)}^{(0)} - \hat B_{(012)}^{(\hat 0)}+\hat B_{(012)} = 1-B_{(012)}+\sum_{a}\bigl(t(\gB_{a})-1\bigr)$.} It can also be interpreted as the number of loops in a connected abstract graph whose vertices are the connected Riemann surfaces in $\gB_{(012)}$ together with the multi-trace vertices in the Feynman graph and whose edges join each multi-trace vertex to all the connected Riemann surfaces it belongs to.

\noindent\emph{Case of connected interaction bubbles}

When the interaction bubbles entering the graph are all connected, $B=1$ and $B^{(0)}=v$, Eq.\ \eqref{hf2} then takes the form
\be\label{hf4} \frac{h}{2} = g+ F_{12}(\gB) - B_{(012)}-B^{(0)}+1\, .\ee
Note that the second term on the right-hand side of \eqref{hf4} is of the form \eqref{Deltadec} for $\gG=\gB$, $p=r$ and the colors $(i_{1}\ldots i_{p})=(3\ldots r+2)$.

\noindent\emph{Case of single-trace interactions}

If the matrix model only includes single-trace interactions, then $t(\gB_{a})=1$  and $B_{(012)}=1$. The formula \eqref{hf2} then yields the usual result
\be\label{hf5} h=2g\, .\ee

\noindent\emph{Second form of the decomposition formula for the index}

The above discussion allows one to rewrite in a very suggestive way the fundamental decomposition formula \eqref{fundid0} for the index.

\begin{proposition} \label{fundidth2} (Second form of the decomposition formula)  The index of a connected $d$-bubble $\gB$, interpreted as the Feynman graph of a tensor model with connected interaction bubbles, is the sum of the quantities $\frac{1}{2}h_{ij}$ that govern the $1/N$ expansions of all the possible $(ij)$ matrix models one can build from the tensor model by singling out any two distinct colors $1\leq i<j\leq d-1$, 
\be\label{fundid1} \ind_{0}\gB = \frac{1}{2}\sum_{i<j}h_{ij}\, .\ee
\end{proposition}
\proof Note that one can always interpret $\gB$ as being a connected Feynman graph in a tensor model with connected interaction bubbles, by setting the rank of the tensor to $R=d-1$ and considering that the connected components of $\gB^{(0)}$ are the interaction bubbles of the model. Previously, in deriving \eqref{hf4}, we singled out the $(12)$ matrix model associated with the colors $1$ and $2$. If we repeat the argument for any choice of colors $i$ and $j$, we find that the amplitude of the $(ij)$ matrix model graph is proportional to $N^{2-h_{ij}}$, where $h_{ij}$ is given by
\be\label{hfij} \frac{h_{ij}}{2} = g(\gB_{(0ij)})+ F_{ij}(\gB) - B_{(0ij)}-B^{(0)}+1\, .\ee
Recall that the term $\smash{F_{ij}(\gB) - B_{(0ij)}-B^{(0)}+1}$ is required because when $\smash{F_{ij}(\gB_{a})>1}$, the $(ij)$ matrix model can be multi-trace, even though the interaction bubbles of the tensor model are themselves connected. Eq.\ \eqref{fundid1} follows from \eqref{fundid0} and \eqref{hfij}.\qed

These considerations suggest to consider an interesting class of tensor models, 
which contain only 
interaction terms $\gB_{a}$ which have $F_{ij}(\gB_{a})=1$ for all choices of colors $i$ and $j$. Let's call interaction bubbles with this property \emph{maximally single-trace} (MST).\footnote{In the mathematical literature, such colorings are referred to as perfect one-factorizations of the graph \cite{perfectone}.} The index 
of a Feynman graph of any model in this particular class is given by the simpler formula
\be \ind_{0}\gB = \sum_{i<j}g(\gB_{(0ij)}) \, . \label{singfaceind}\ee
This family of interactions will be further studied in Sec.\ \ref{MSTSec} below and App.\ \ref{appendixA}.

\subsubsection{Factor of \texorpdfstring{$D$}{D}}

Let us now evaluate the factor of $D$ in \eqref{Feyn1}; we read off
\begin{multline}\label{ellf1} \frac{\ell}{r+1} = r+2g+2(1-B_{(012)}) +\sum_{a}t(\gB_{a})-(r+2)v+r p\\
+\sum_{a}c(\gB_{a})-\varphi-\frac{2}{(r+1)!}\sum_{a}\deg\gB_{a}\, ,
\end{multline}
where we used Eq.~\eqref{hf2}. Using the relations \eqref{fcrel1} together with the second equation in \eqref{B012genus}, as well as
\be\label{fcrel2} 
\sum_{a}c(\gB_{a}) = B^{(0)}\, , \quad\varphi = \sum_{I}F_{0I}\, ,\quad
\sum_{a}\deg\gB_{a} =\deg\gB^{(0)}\, ,\ee
we can rewrite $\ell$ as
\be\label{ellf2} \frac{\ell}{r+1} = r+2+ B^{(0)}-(r+2) v+\frac{1}{2}(r+1)V(\gB)-\sum_{i}F_{0i}-\frac{2}{(r+1)!}\deg\gB^{(0)}\, .\ee
We then apply the formula \eqref{degreeformula} for the degree to the bubbles $\gB^{(0)}$ and $\gB$,
\begin{align}\label{deg1f} & (r+1)B^{(0)}-\frac{2}{r!}\deg\gB^{(0)}=F(\gB)-\sum_{i}F_{0i}(\gB)-\frac{1}{4}r(r+1)V(\gB)\, ,\\\label{deg2f} &
(r+2)B-\frac{2}{(r+1)!}\deg\gB = F(\gB)-\frac{1}{4}(r+1)(r+2)V(\gB)\, .
\end{align}
Subtracting these two equations, plugging the result into \eqref{ellf2} and using the definition \eqref{inddef} of the index finally yields
\be\label{ellindex}\ell= 2\ind_{0}\gB + (r+1)(r+2)\Bigl[\sum_{a}\bigl( c(\gB_{a})-1\bigr) - B + 1\Bigr]\, .\ee
The right-hand side of \eqref{ellindex} is the sum of two positive terms. The positivity of the second term follows straightforwardly from the connectivity inequality \eqref{ineq} or from its interpretation as the number of loops in a suitably constructed abstract graph, in full parallel with the discussion following Eq.\ \eqref{posterms1}. Moreover, \eqref{ind0N} implies that $\ell\in\mathbb N$. Finally, in the special case of connected interaction bubbles, one has $c(\gB_{a})=1$, $B=1$ and thus
\be\label{ellindex3}\ell= 2\ind_{0}\gB\, .\ee
%


\vfill\eject

\subsubsection{Form of the expansions}

\noindent\emph{Generalities}

The above results, most crucially the fact that $\ell\geq 0$, imply that the free energy $F$, or the correlation functions which are obtained from the free energy by taking partial derivatives with respect to the couplings, have well-defined large $N$ and large $D$ expansions for the scalings \eqref{newcouplings}. More precisely, we first expand at large $N$,
\be\label{Fexp} F = \sum_{h\in\mathbb N}F_{h}\, N^{2-h}\, ,\ee
where the $F_{h}$ are $N$-independent coefficients. We can then expand each $F_{h}$ at large $D$,
\be\label{Fhexp} F_{h}=\sum_{\ell\in\mathbb N}F_{h,\ell}\, D^{r+h-\frac{\ell}{r+1}}\, .\ee
Note that, since $\ell$ is an integer, the natural expansion parameter is $1/D^{\frac{1}{r+1}}$. It is important to understand that the limits $N\rightarrow\infty$ and $D\rightarrow\infty$ do not commute, as in \cite{ferra1}. Indeed, even if the power of $D$ is bounded above at fixed $h$, it can grow linearly with $h$ at fixed $N$. This means that the limit $D\rightarrow\infty$ at fixed $N$ does not exist. One must always consider first $N\rightarrow\infty$ and second $D\rightarrow\infty$, at each order in the $1/N$ expansion.

It is also possible to set $D=N$ and take the $N\rightarrow\infty$ limit, which yields
\be\label{Ftensorexp} F = \sum_{\ell\in\mathbb N}\hat F_{\ell}\, N^{R-\frac{\ell}{r+1}}\, .\ee
This is a new expansion for general tensor models. Compared to the expansion studied in \cite{BGR}, the expansion parameter is $1/N^{\frac{1}{r+1}}$ instead of $1/N$ and the expansion is governed by the index of the Feynman graphs instead of their degree.

\noindent\emph{Leading order graphs}

The leading order graphs satisfy
\begin{align}\label{leadingcond1} &\sum_{a}\bigl( c(\gB_{a})-1\bigr) =B-1,\\
\label{leadingcond2}  &\ind_{0}\gB = 0\, .
\end{align}
The first condition \eqref{leadingcond1} is automatically satisfied in models with connected interaction bubbles. Otherwise, it implies that $\gB$ is ``maximally disconnected'' for a given set of vertices, taking into account the fact that the Feynman graph itself is connected. The second condition says that the connected components of the leading graphs must all have index zero, i.e.\ must be generalized melons, as defined in Sec.\ \ref{gmSec}. As we have already mentioned in that section, the class of generalized melons is much larger than the class of ordinary melons that dominate the large $N$ limit in the BGR scaling. In particular, in our new scaling, non-melonic interaction vertices can contribute at leading order. As mentioned earlier, the sum over generalized melons may better capture interesting physics than the sum over ordinary melons.

\noindent\emph{Upper bound on the power of $D$ at fixed $h$}

From \eqref{Fhexp} it is clear that, at fixed $h$, the highest possible power of $D$ in a Feynman graph is $r+h$. For single-trace models, this is $r+2g$. This upper bound can be easily improved as follows. From the decomposition \eqref{fundid0} and the identification \eqref{ggrel}, we get
\be\label{indgt} \ind_{0}\gB\geq g + F_{12}-B_{(012)}-B^{(0)}+B\, .\ee
Using \eqref{hf2} and the first equalities in \eqref{fcrel1} and \eqref{fcrel2}, we obtain
\be\label{indgt2} \ind_{0}\gB\geq \frac{h}{2}-\sum_{a}\bigl(c(\gB_{a})-1\bigr) + B -1\, .\ee
Together with \eqref{ellindex}, this yields
\be\label{ellgt} \ell\geq h+r(r+3)\Bigl[\sum_{a}\bigl( c(\gB_{a})-1\bigr) - B + 1\Bigr]\, .\ee
Using \eqref{Fhexp}, we thus see that the highest possible power of $D$ is actually
\be\label{Dpowermax} r+\frac{r}{r+1}h - \frac{r(r+3)}{r+1}\Bigl[\sum_{a}\bigl( c(\gB_{a})-1\bigr) - B + 1\Bigr]\, .\ee
For single-trace models, this reduces to $r+\frac{2r}{r+1}g$, generalizing the bound $1+g$ obtained in the case $r=1$ in \cite{ferra1}.

\subsubsection{Other scalings and expansions}

Let us finally mention two other natural scalings that yield non-trivial large $D$ and/or large $N$ expansions, albeit keeping less diagrams.

\noindent\emph{BGR scaling}

This scaling  is defined by \eqref{BGRcouplings}. Straightforward modifications of the derivations presented above show that a Feynman diagram is proportional to $N^{2-h}D^{r+h-L}$, where $h$ is given by \eqref{hf2} and $L$ by
\be\label{Lform} L = \frac{2}{(r+1)!}\deg\gB + (r+2)\Bigl[\sum_{a}\bigl( c(\gB_{a})-1\bigr) - B + 1\Bigr]\, .\ee
We find that $L\in\mathbb N$, using in particular \eqref{degreeN}. We thus get an expansion in powers of $1/D$ governed by the degree. 

Using the inequalities \eqref{degineq} successively for the colors 3 to $R$, together with $\deg\gB_{(012)}=g$, we find that
\be\label{deggt} \deg\gB\geq\frac{1}{2}(r+2)!\, g\, .\ee
For single-trace models, the highest power of $D$ that can appear at a given genus $g$ is thus $(1-g)r$, generalizing the bound $1-g$ found for $r=1$ in \cite{ferra1}. This result is a kind of non-renormalization theorem, showing that, at a given order $\smash{D^{r-\hat L}}$ in the $1/D$ expansion, only graphs of genus $\smash{0\leq g\leq\frac{\hat L}{r}}$ can contribute. This bound also shows that the power of $D$ is bounded above by $r$ independently of the genus of the graph: unlike with the enhanced scaling \eqref{newcouplings}, the large $N$ and large $D$ limits commute in the BGR scaling.

\noindent\emph{Splitted scaling}

Another natural procedure is to define a scaling by treating the matrix and tensor parts of our variables separately.\footnote{Even more generally, we could consider splitted scalings for which we separate the $R$ indices of the tensor into several groups, $R=r_{1}+\cdots +r_{s}$.} We always use the standard 't~Hooft's scaling for the matrix part, which means that the couplings $\tau_{a}$ in \eqref{genaction} do not depend on $N$. For $r=1$, we decide that they do not depend on $D$ either. This corresponds to the usual large $D$ scaling of vector models. For $r=2$, which is a bi-matrix model, we use an action of the form
\be\label{act2} S = ND\Bigl( \tr X_{\mu\nu} X^{T}_{\mu\nu} + \sum_{a}N^{1-t(\gB_{a})}D^{1-\tilde t(\gB_{a})}\kappa_{a}I_{\gB_{a}}(X)\Bigr)\, ,\ee
where we have defined $\tilde t(\gB_{a})=F_{34}(\gB_{a})$ and we keep the couplings $\kappa_{a}$ fixed. For $r\geq 3$, we choose to use our enhanced scaling for the tensor part of the matrix variable, which amounts to keeping the $\kappa_{a}$ defined by
\be\label{act3} S = ND^{r-1}\Bigl( \tr X_{\mu_{1}\cdots\mu_{r}} X_{\mu_{1}\cdots\mu_{r}} + \sum_{a}N^{1-t(\gB_{a})}D^{1-c(\gB_{a}^{(12)})+\frac{2}{(r-1)!}\deg\gB_{a}^{(12)}}\kappa_{a}I_{\gB_{a}}(X)\Bigr)\ee
fixed. It is easy to check that these scalings are less optimal than \eqref{newcouplings}, in the sense that the ratios $\kappa_{a}/\la_{a}$ are always proportional to a positive power of $D$. The large $N$ and large $D$ limits always commute in the splitted scalings, because the highest power of $D$ of any Feynman diagram is $D^{r}$.

%

%

%
\subsection{\label{MSTSec}Optimal scalings and MST interactions}
\begin{figure}
\begin{center}
{\includegraphics[width=6in]{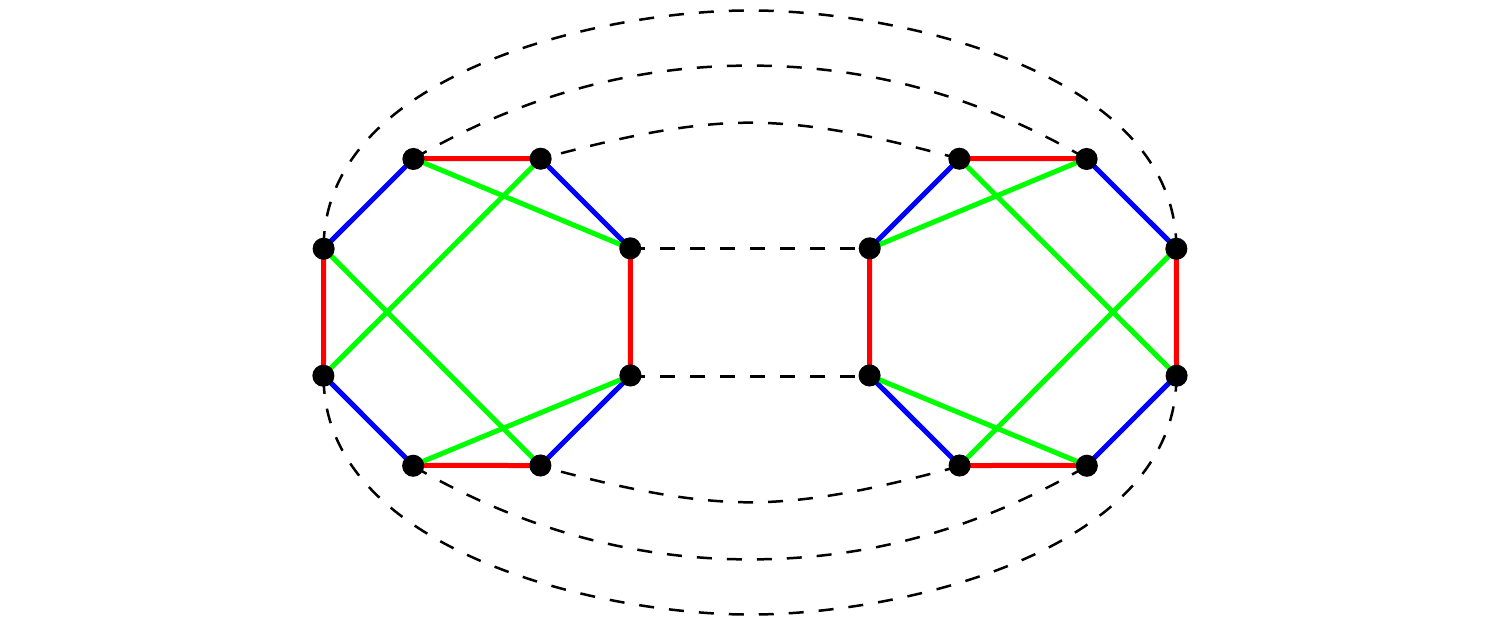}}
\end{center}
\caption{A mirror melon built from two identical bubbles.}
\label{figureGraphMirror}
\end{figure}

\subsubsection{\label{MirrorSec}Mirror melons}

Consider any interaction bubble $\gB$ with \emph{labeled vertices}. If we join vertices 
\emph{with identical labels} in two copies of $\gB$ by an edge of color 0, we obtain what we call the \emph{elementary mirror melon} $\gM_0(\gB)$ associated to $\gB$. It is convenient to picture the second bubble as a mirror reverse image of the first, as depicted in Fig.\ \ref{figureGraphMirror}; hence the name. In analogy with Sec.\
\ref{gmSec}, we can then construct an infinite family of graphs called the mirror melons associated to $\gB$. Indeed, for any edge $\ell$ of color 0 in $\gM_0(\gB)$, 
there is a 2-point graph $\gM^\ell_0(\gB)$ obtained by cutting open the edge $\ell$. Inserting the graph $\gM^\ell_0(\gB)$ on a edge of color $0$ is then called a mirror melonic move and iterating this operation yields the infinite family of mirror melons 
associated to $\gB$.

Note that there is no reason that generalized melons should coincide with mirror melons.  In the next subsection, we prove that mirror melons are generalized melons if and only if  $\gB$ is MST.

%
%
%

%
\subsubsection{\label{MSTSec2}The case of MST interactions}

To prove that a scaling is optimal for a given interaction term and to identify the leading graphs are two different
and difficult problems in general. We now give a simple criterion which allows to conclude for the first problem.
\begin{lemma}\label{optimallem} If a given interaction bubble can contribute at leading order (i.e.\ can be part of a generalized melon), then our scaling \eqref{newcouplings} is optimal for the associated coupling.
\end{lemma}
\proof If a given interaction bubble $\gB_{a}$, with associated coupling $\tau_{a}$ that scales as in \eqref{newcouplings}, is part of a generalized melon, then we can use this generalized melon and the generalized melonic moves described at the end of Sec.\ \ref{gmSec} to build new generalized melons containing an arbitrary number of bubbles $\gB_{a}$. If we enhance further the large $D$ scaling of $\tau_{a}$, we thus get planar graphs that can be proportional to an arbitrary high power of $D$. The $D\rightarrow\infty$ limit is thus no longer well-defined.\qed

As a simple application, we can easily study the case of all MST bubbles. Let $\gB_{\text{MST}}$ be a MST bubble with $R=r+2$ colors and $V(\gB_{\text{MST}})=V$ vertices. Such a bubble is automatically connected and has $\frac{1}{2}R(R-1)$ faces. Eq.\ \eqref{degreeformula} then yields $\deg\gB_{\text{MST}} = \frac{1}{8}(R-2)(R-1)!(V-2)$ or, equivalently, a scaling \eqref{newcouplings} of the form
\be\label{scaleMST}\tau_{\text{MST}}=D^{\frac{1}{4}r(V-2)}\la_{\text{MST}}\, .\ee
\begin{proposition}\label{MSToptprop} The scaling \eqref{scaleMST} is optimal for all MST interaction terms. 
\end{proposition}
\proof Consider an MST bubble $\gB_{\text{MST}}$ and its associated elementary mirror melon $\gM_0(\gB)$
as in Fig.\ \ref{figureGraphMirror}. 
By construction, all the $(0i)$-faces of $\gB$ contain one edge of color $i$ in each MST bubble. Their total number thus matches with the number of edges in $\gB_{\text{MST}}$, $\sum_{i}F_{0i}(\gB)=E(\gB_{\text{MST}})=\frac{1}{2}RV$. Moreover, using the MST property, $\smash{\sum_{i<j}F_{ij}(\gB)=2\sum_{i<j}F_{ij}(\gB_{\text{MST}}) = R(R-1)}$. We also trivially have $B=1$ and $B^{(0)}=2$. Eq.\ \eqref{indexplicit} then yields $\ind_{0}\gB=0$ and we conclude using Lemma \ref{optimallem}.\qed

Note that it is easy to check that the elementary mirror melon yields a generalized melon if and only if the interaction bubbles are MST.

A natural but much more difficult question to ask is whether the mirror melons yield all the possible generalized melons. The aim of the next section will be to prove that this is indeed true in the particular case of the complete interaction on $R+1$ vertices, when $R \ge 3$ is a \emph{prime number}.

\section{\label{applicationsSec}The case of the complete interaction}

We are now going to study in detail the model based on the complete interaction of odd rank $R$. The case $R=3$ was solved by Carrozza and Tanasa in \cite{CarrozzaTanasa}. Interestingly, the cases $R>3$ turn out to be qualitatively different and their analysis requires to introduce several new ingredients. Our main result is a full classification of the leading graphs in our new scaling, under the condition that $R$ is a prime number. One can then construct quantum mechanical models akin to the SYK model with $q=(R+1)$-fold random interaction \cite{maldastan}. The classification in the cases where $R$ is not prime require further non-trivial extensions of the formalism and are beyond the scope of the present paper.

\noindent\emph{Notations}\\ Labeled vertices belonging to an interaction bubble are denoted in brackets, like $[\nu_{1}], [\nu_{2}]$, etc. When several interaction bubbles are present, an upper index may be added to distinguish between the bubbles, like $[\nu_{1}]^{I}$, $[\nu_{1}]^{II}$, etc. A path going successively through the vertices $[\nu_{1}],[\nu_{2}],\ldots,[\nu_{q}]$ is denoted as $[\nu_{1}][\nu_{2}]\cdots[\nu_{q}]$. This is unambiguous inside interaction bubbles, which have at most one edge joining two given vertices. In other cases, a possible ambiguity is waived by specifying the edge colors. The path is oriented if we distinguish between $[\nu_{1}][\nu_{2}]\cdots[\nu_{q}]$ and $[\nu_{q}][\nu_{q-1}]\cdots [\nu_{1}]$. A path $[\nu_{1}][\nu_{2}]$ is an edge. 

Equality in $\mathbb Z/R\mathbb Z$, i.e.\ equality modulo $R$, is denoted as $\equiv$. When $R$ is prime, $\mathbb Z/R\mathbb Z$ is a field in the algebraic sense and the inverse of an element $x$ is denoted by $x^{-1}$; for example, $2^{-1}\equiv\frac{1}{2}(R+1)$.

\subsection{Properties of the complete graph and its edge-coloring}

\subsubsection{\label{compdefSec}Definition of the complete interaction bubble}

\begin{figure}
\centerline{\includegraphics[width=6in]{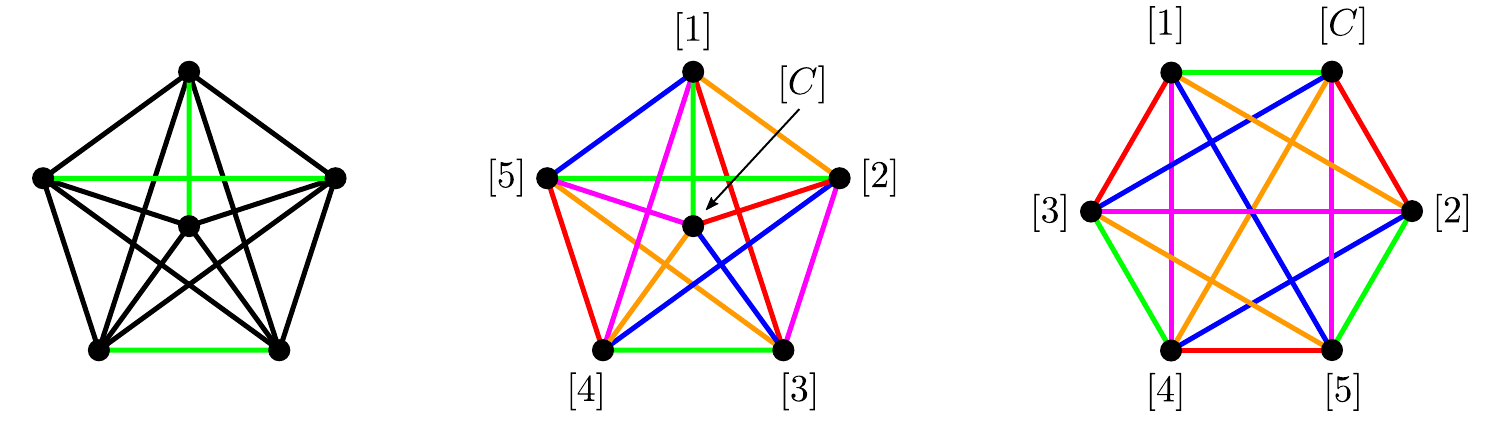}}
\caption{Edge-coloring for the complete graph $K_6$. Left: rule for the coloring of the edges of a particular color, here green. Center: full edge-coloring and standard vertex labeling, here $1=\text{green}$, $2=\text{red}$, $3=\text{blue}$, $4=\text{orange}$ and $5=\text{purple}$. Right: the equivalent $(\text{green},\text{red})$-polygonal representation in the shape of a six-sided polygon whose boundary is the face of colors green and red. This polygonal representation is natural when $R$ is prime.}\label{figureA}
\end{figure}

For $R$ odd, the complete graph  $K_{R+1}$ with $R+1$ vertices and $\frac{1}{2}R(R+1)$ edges is edge-colorable with $R$ colors. This is related to the scheduling of a Round-Robin Tournament as pointed out in \cite{Narayan:2017qtw}. The explicit $R$-regular edge-coloring that we shall use can be described as follows \cite{soifer}. We consider a regular $R$-sided polygon plus its center. The center is labeled as $[C]$ and the vertices of the polygon are cyclically numbered as $[1]$ to $[R]$. For each color $1 \le i \le R$, draw an edge of color $i$ from the center $[C]$ to the vertex $[i]$ of the polygon. Then, use the same color for all edges between polygon vertices that are perpendicular to the edge $[C][i]$. If we identify the polygon vertices with ${\mathbb Z } / R {\mathbb Z }$, it means that the polygon vertices $[n]$ and $[n'] \ne [n]$ are joined by an edge of color $i$ if and only if $n+n'\equiv 2i$. Equivalently, the edges of color $i$ join the polygon vertices $[i+p]$ and $[i-p]$ for all $1\leq p\leq \frac{1}{2}(R-1)$. The $R$-bubble obtained in this way will be denoted as $\gK_{R+1}$ and the above color and vertex labeling will be called the ``standard'' coloring. The construction is illustrated in Fig.\ \ref{figureA} for $\mathcal K_6$. 

\begin{figure}
\centerline{\includegraphics[width=2.8in]{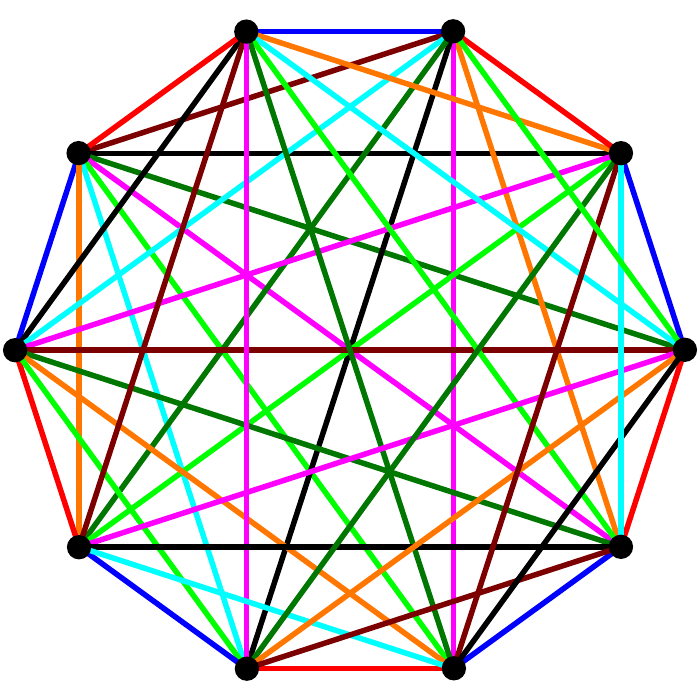}}
\caption{MST coloring of $K_{10}$.}\label{nonstandardK10}
\end{figure}

We shall say that two edge-colorings for the complete graph are \emph{equivalent} if there exist a permutation $\sigma$ of the $R+1$ vertices and a permutation $\tau$ of the $R$ colors that change one edge-coloring into the other. It is easy to check directly that all the possible colorings for $R=3$ and $R=5$ are equivalent to the standard one. More generally, the number of non-equivalent edge-colorings for the complete graph is counted by the sequence A000474 in OEIS \cite{OEIS}.\footnote{We would like to thank Fidel Schaposnik for pointing this out to us.} For example, there are six non-equivalent edge-colorings for $K_8$, 396 for $K_{10}$, etc. If we also impose the MST condition, there remains only one possibility for $K_{8}$, which is the standard coloring $\mathcal K_{8}$, and also one possibility for $K_{10}$, which is \emph{not} the standard coloring (we shall demonstrate below that the standard colorings are MST if and only if $R$ is a prime number; $\mathcal K_{8}$ is thus MST, but $\mathcal K_{10}$ is not). The non-standard MST coloring for $K_{10}$ is depicted in Fig.\ \ref{nonstandardK10}.\footnote{We thank Fidel Schaposnik for providing this example} 

In the following, we only focus on the complete interaction bubble with the standard coloring. All our results will strongly depend on this choice and it is an open question as to whether similar results can be derived for edge-colorings that are not equivalent to the standard one. For instance, we do not know the classification of the generalized melons in the case of the non-standard MST coloring of $K_{10}$ depicted in Fig.\ \ref{nonstandardK10}.


\subsubsection{The MST condition}
From now on we consider only the standard coloring.
The cases $R$ prime and $R$ not prime are qualitatively different, as indicated by the following proposition.

\begin{proposition} \label{prime}
The $R$-bubble $\gK_{R+1}$ is maximally single-trace (MST) if and only if $R$ is a prime number.
\end{proposition}

\proof

For any pair of two colors $(i,j)$, let us consider the $(ij)$-face that goes through the center vertex $[C]$. Let us denote its length as $2q$, which is the even integer defined to be the number of its edges of colors $i$ and $j$.  Using the rule for the edge-coloring of the complete bubble, we can explicitly write this face, starting from the center vertex $[C]$ with the edge of color $j$, as $[C][k_1\equiv j][k_2]\cdots [k_{2q-1}\equiv i][C]$ and check inductively that $k_{2p}\equiv 2pi-(2p-1)j$ and $k_{2p+1}\equiv (2p+1)j - 2pi$ for $1\leq p\leq q-1$. Therefore, $k_{2q-1}\equiv i$ is equivalent to $(2q-1)(j-i) \equiv 0$. 

If $R$ is a prime number, this implies $2q-1 \equiv 0$ because $j-i \not\equiv 0$. The length is the smallest possible solution, that is, $2q=R+1$. Our $(ij)$-face must then visit all the vertices of $\gK_{R+1}$ and is thus unique: $F_{ij}=1$ for all pairs $(i,j)$ and the bubble is MST. 

If $R$ is not prime, write $R=R_{1}R_{2}$ where $R_{1}$ and $R_{2}$ are odd integers with $1<R_{1}<R$ and $1<R_{2}<R$. Moreover, set $j-i=R_{2}$. The smallest solution to $(2q-1)(j-i) \equiv 0$ is then $2q=R_{1}+1<R+1$. This implies that there are vertices in $\gK_{R+1}$ that are not visited by our $(ij)$-face, which therefore cannot be unique: $F_{ij}>1$ and the bubble is not MST.\qed

%

For instance, when $R=9$, there are two $(14)$-faces, namely $[C][1][7][4][C]$ of length four going through the center and $[2][6][5][3][8][9][2]$ of length six.

The bubbles $\gK_{R+1}$ with $R$ prime will be called \emph{prime-complete}. These bubbles have convenient \emph{$(ij)$-polygonal} representations in the shape of an $(R+1)$-sided polygon whose boundary is the unique $(ij)$-face. This is illustrated on the right of Fig. \ref{figureA} for $\mathcal K_6$.

A simple application of Proposition \ref{prime} is the computation of the degree of the prime-complete bubbles. Indeed, since we know that there is exactly one face per pair of colors, the total number of faces is simply $F(\gK_{R+1}) = \frac{1}{2}R(R-1)$. Together with $c(\gK_{R+1})=1$ and $V(\gK_{R+1}) = R+1$, Eq.\ \eqref{degreeformula} yields
\begin{equation}  
\frac{2}{(R-1)!}\deg \gK_{R+1} = \frac{1}{4} (R-1)(R-2)\, .
\label{degreeint}
\end{equation}
In contrast, for $R$ not a prime, we have by Proposition \ref{prime} that $F(\gK_{R+1}) > \frac{1}{2}R(R-1)$ so that
\begin{equation}  
\frac{2}{(R-1)!}\deg \gK_{R+1} < \frac{1}{4} (R-1)(R-2)\, .
\label{degreeintnotprime}
\end{equation}

\subsubsection{\label{disbubbleSec}Distinguishing edges and vertices}

It is very easy to check that all the edges of a given color in $\gK_{4}$ are equivalent and that all the vertices of $\gK_{4}$ are equivalent too. As we now explain, the situation is drastically different at higher rank. It turns out that the $\frac{1}{2}(R+1)$ edges of any given color in the prime-complete bubble $\gK_{R+1}$ are all inequivalent, for all primes $R>3$. The same is true for the $R+1$ vertices of the bubble as well. This major difference between $R=3$ and $R>3$ goes a long way in explaining why a new proof of the classification theorem for the leading graphs must be given.

An elegant way to distinguish between edges of a given color is as follows. Consider an oriented edge $[\nu_{1}][\nu_{2}]$ of color $k$. For any ordered pair of distinct colors $(i,j)$, there exists a unique $(ij)$-path, i.e.\ a path of alternating colors $i$ and $j$, that starts at $[\nu_{1}]$ with an edge of color $i$ and ends at $[\nu_{2}]$. The existence of this path is ensured by the fact that the unique $(ij)$-face visits all the vertices of the prime-complete bubble. If $\ell$ is the length of the path, defined to be the number of its edges of colors $i$ and $j$, we then say that the ordered pair of colors $(i,j)$ \emph{indexes the oriented edge $[\nu_{1}][\nu_{2}]$ at length $\ell$}. Unoriented edges can also be indexed by unordered pairs in an obvious way. The indexing enjoys the following simple properties.

\begin{lemma}\label{simpleindex} The edge $[\nu_{1}][\nu_{2}]$ is indexed by $(i,j)$ at length $\ell$ if and only if is it indexed by $(j,i)$ at length $\ell'=R+1-\ell$. The edge $[\nu_{1}][\nu_{2}]$ is indexed by $(i,j)$ at even length $\ell$ if and only if the edge $[\nu_{2}][\nu_{1}]$ is indexed by $(j,i)$ at even length $\ell$. The edge $[\nu_{1}][\nu_{2}]$ is indexed by $(i,j)$ at odd length $\ell$ if and only if the edge $[\nu_{2}][\nu_{1}]$ is indexed by $(i,j)$ at odd length $\ell$.\end{lemma}

\proof Trivial by following the $(ij)$-face.\qed

We say that two oriented edges of the same color are \emph{weakly equivalent} if they are indexed by the same set of pairs of colors at length two. Otherwise, they are \emph{strongly inequivalent}. 


At length one, any edge of color $k$ is obviously indexed by the $R-1$ pairs $(k,i)$ for all $i\not = k$. For $R=3$, any edge of color $k$ is indexed at length two by the two ordered pairs $(i,j)$ and $(j,i)$ of complementary colors. For $R>3$, the computation of the pairs of colors indexing an arbitrary edge at any length $\ell\geq 2$ is a straightforward exercice. The results at lengths two and three are summarized by the following lemma.

\begin{lemma} \label{lengthtwolemma} Consider the bubble $\mathcal K_{R+1}$ for $R$ prime and $R>3$. We use the standard vertex labeling and we name the colors by integers modulo $R$.

The edge $[C][k]$, of color $k$, is indexed at length two by the pairs of colors $(i,j)$ with $j\equiv \frac{1}{2}(R+1)(k+i)$, for all $i\not\equiv k$. This yields $R-1$ distinct pairs. It is indexed at length three by the pairs of colors $(i,j)$ with $j\equiv \frac{1}{2}(R+1)(3i-k)$, for all $i\not\equiv k$. This yields $R-1$ distinct ordered pairs.

The edge $[k-p][k+p]$, for any $1\leq p\leq \frac{1}{2}(R-1)$, of color $k$, is indexed at length two by the pairs of colors $(k-p,k+p)$ and $(i,i+p)$ for all $i$ different from $k$ and $k-p$. This yields a total of $R-1$ distinct pairs. It is indexed at length three by the pairs of colors $(k-p,k-3p)$, $(k+p, k+3p)$ and $(i,2i-k)$ for all $i$ different from $k$, $k-p$ and $k+p$. This yields $R-1$ distinct ordered pairs.
\end{lemma} 
%



\proof
The results follow from a direct computation using the rule of the edge-coloring. For example, the edge of color $i\not\equiv k$ attached to $[C]$ is $[C][i]$. The edge $[i][k]$ is then of color $j$ with $i+k\equiv 2j$, which proves the first statement since $2^{-1}\equiv\frac{1}{2}(R+1)$ in $\mathbb Z/R\mathbb Z$. At length three, one again considers the edge $[C][i]$ of any color $i\not\equiv k$ and the edge of color $i$ starting from $[k]$, which is $[k][2i-k]$. The color of the edge $[i][2i-k]$ is then $j\equiv 2^{-1}(3i-k)$, which proves the second statement. The results for the edges $[k-p][k+p]$ follow from a similar analysis.\qed

From Lemma \ref{lengthtwolemma}, we can easily prove two simple results which will be useful later.

\begin{proposition} \label{inequivalent}
For $R$ prime and $R>3$, two distinct oriented edges of the same color in $\gK_{R+1}$ are always strongly inequivalent. Equivalently, two weakly equivalent edges necessarily coincide. 
\end{proposition} 

\proof Colors are defined modulo $R$. Lemma \ref{simpleindex} implies that if $(i,j)$ indexes an edge $[\nu_{1}][\nu_{2}]$ at length two, then $(j,i)$ indexes $[\nu_{2}][\nu_{1}]$ at lenght two but does not index $[\nu_{1}][\nu_{2}]$ at length two as soon as $R>3$. Thus $[\nu_{1}][\nu_{2}]$ and $[\nu_{2}][\nu_{1}]$ are strongly inequivalent. Moreover, using Lemma \ref{lengthtwolemma}, it is straightforward to check that, for any $k$, the pair of colors $(k-p,k+p)$ indexes the edge $[k-p][k+p]$ at length two for all $p\not\equiv 0$ but does not index the edge $[C][k]$ at length two; and for all $p'\not\equiv p$, $p\not\equiv 0$, $p'\not\equiv 0$, the pair of colors $(k+\frac{1}{2}(R+1)(p'-p),k+\frac{1}{2}(R+1)(p'+p))$ indexes the edge $[k-p][k+p]$ at length two but does not index the edge $[k-p'][k+p']$ at length two.\qed

We thus see that two distinct oriented edges of a given color in the prime-complete bubble can be unambiguously distinguished from one another using the coloring of the graph. The same is then automatically true for the vertices, since two distinct vertices $[\nu_{1}]$ and $[\nu_{2}]$ are the endpoints of two distinct oriented edges $[\nu_{1}][\nu_{2}]$ and $[\nu_{2}][\nu_{1}]$.

\vfill\eject

\begin{lemma} \label{lengththree}
For $R$ prime and $R>3$, choose two distinct unoriented edges $[\nu_{1}][\nu_{2}]$ and $[\nu_{3}][\nu_{4}]$ (i.e.\ such that $[\nu_{1}][\nu_{2}]\not = [\nu_{3}][\nu_{4}]$ and $[\nu_{1}][\nu_{2}]\not = [\nu_{4}][\nu_{3}]$) of the same color $k$. One can then always find a pair of colors $(i,j)$ indexing one of the edge at length two and the other at length three. Note that, of course, $i\not = k$ and $j\not = k$.
\end{lemma} 
\proof
If the two distinct edges of color $k$ are $[C][k]$ and $[k-p][k+p]$ for some $1\leq p\leq \frac{1}{2}(R-1)$, one considers the pair of colors $(k+2p,k+3p)$. Using Lemmas \ref{simpleindex} and \ref{lengthtwolemma}, it is straightforward to check that it indexes $[k-p][k+p]$ at length two and both $[C][k]$ and $[k][C]$ at length three. Similarly, $(k-2p,k-3p)$ indexes $[k+p][k-p]$ at length two and both $[C][k]$ and $[k][C]$ at length three. If the two distinct edges of color $k$ are $[k-p][k+p]$ and $[k-p'][k+p']$ for some $1\leq p,p'\leq \frac{1}{2}(R-1)$, $p\not = p'$, one considers the pair of colors $(k+p',k+2p')$. Using again Lemmas \ref{simpleindex} and \ref{lengthtwolemma}, we see that it indexes $[k-p'][k+p']$ at length two and both $[k-p][k+p]$ and $[k+p][k-p]$ at length three. Similarly, $(k-p',k-2p')$ indexes $[k+p'][k-p']$ at length two and both $[k-p][k+p]$ and $[k+p][k-p]$ at length three.\qed

\subsection{Action and index}

The interaction term associated with the complete bubble $\gK_{R+1}$ is given explicitly by
\be\label{uncoloredint}
I_{\gK_{R+1}}(T) = \prod_{\nu=0}^{R} T_{a_{\nu,1}  \cdots \,  a_{\nu,R}} \, ,
\ee
where $T$ is a real tensor of rank $R$. In the expression \eqref{uncoloredint}, we set $a_{0,i}=a_{i,i}$ and $a_{n,i}=a_{n',i}$ for $1\leq n,n'\leq R$, $n'\not = n$ and $n+n' \equiv 2i$, so that we reproduce the edge-coloring of $K_{R+1}$ explained above. We also sum over repeated indices. We want to study the model with action
\be \label{actionGenCT} S=N^{R-1} \Bigl( \frac{1}{2} T_{a_1 \cdots a_R} T_{a_1 \cdots a_R} +  N^{\frac{1}{4}(R-1)(R-2)}  \la\, I_{\gK_{R+1}}(T) \Bigr) \, , \ee
for $R$ prime.\footnote{As usual, instead of a zero-dimensional action, we could consider quantum mechanical or field-theoretic generalizations. Our subsequent discussion would remain unchanged.} The particular power of $N$ in front of $\la$ in \eqref{actionGenCT} has been chosen to match our new enhanced scaling, consistently with \eqref{genaction} and \eqref{newcouplings} at $N=D$ and with the formula \eqref{degreeint} for the degree of the prime-complete bubble.

As explained in Sec.\ \ref{largeNDSec}, when one takes the large $N$ limit at fixed $\la$, one gets a well-defined expansion in powers of $1/N^{\frac{1}{R-1}}$. The equations \eqref{ellindex3} and \eqref{Ftensorexp} show that connected vacuum Feynman graphs $\gB$ contributing at a given order $\smash{N^{R-\frac{\ell}{R-1}}}$ have a fixed index $\smash{\ind_{0}\gB=\frac{\ell}{2}}$. An elegant formula for the index in our model is given by \eqref{singfaceind}, since the prime-complete interaction is maximally single-trace. A more explicit formula can be obtained from \eqref{indexplicit}. Since presently the Feynman graph vertices are all prime-complete bubbles, we have $B=1$, $B^{(0)}=v$, $V=(R+1)v$ and $\sum_{i<j}F_{ij}=\frac{1}{2}R(R-1)v$, which yields
\begin{equation}\label{indexmodel}
\ind_{0}\gB =\frac{1}{2}R(R-1) + \frac{1}{8}R(R-1)^2 v -\frac{1}{2}(R-1)\sum_i F_{0i} \, .
\end{equation}
The graphs dominating the large $N$ expansion are the generalized melons, which are, by definition, of index zero. The generalized melons of our model will be called the \emph{prime-complete generalized melons}, or PCGMs for short. They maximize the total number of $(0i)$-faces for a fixed number of vertices. The condition for a Feynman graph to be a PCGM is equivalent to  
\begin{equation}
\sum_i F_{0i} = \frac{1}{4}R(R-1)v + R \, ,
\label{leadinggraphs}
\end{equation}
which, using \eqref{singfaceind}, is itself equivalent to
\be\label{leadg} g(\gB_{(0ij)}) = 0\ \text{for all pairs of distinct colors $(i,j)$.}\ee
In the next subsection, we are going to solve explicitly these conditions and provide a full description of all the PCGMs.

\subsection{The classification theorem}

\subsubsection{Useful tools}

The simple lemma below will be used repeatedly in the following.
\begin{lemma}\label{cyclelemma} A planar 3-bubble cannot have cycles of odd length. 
\end{lemma}
\proof This is a direct consequence of a more general result, explained in Sec.\ \ref{bubjackSec}, which states that the underlying graph of an orientable bubble is bipartite, together with the well-known facts that a planar graph is orientable and that a graph is bipartite if and only if it does not contain cycles of odd length.\qed

We shall also need standard results on the deletion of edges and vertices from a planar ribbon graph. The edge deletion is defined in the trivial way, maintaining the cyclic ordering of the remaining edges around vertices. The vertex deletion is defined only for vertices of valency two. It simply amounts to replacing the two ribbons attached to the vertex by a unique ribbon, twisted if precisely one of the two original ribbons is twisted or untwisted otherwise. It is useful to introduce the following terminology \cite{book}: an edge of a ribbon graph is called \emph{regular} if it belongs to two distinct faces and it is called \emph{singular} otherwise; in other words, the borders of the ribbon associated with a regular edge are on two distinct faces, whereas they are on the same face in the case of a singular edge.
\begin{lemma}\label{oplemma} i) (Edge deletion) If one deletes a regular edge from a connected planar ribbon graph, one gets another connected planar ribbon graph. If one deletes a singular edge from a connected planar ribbon graph, one gets a ribbon graph with two planar connected components.\\
ii) (Vertex deletion) If one deletes a vertex of valency two from a connected planar ribbon graph, one gets a connected planar ribbon graph.
\end{lemma}
The claims i) are special cases of standard results on edge deletions for ribbon graphs of arbitrary genus. The proof, which is elementary, will not be included here, see e.g.\ \cite{book}. The claim ii) is trivial to check.

\subsubsection{Results on faces}
\begin{lemma}\label{revisitlemma} A $(0i)$-face in a PCGM cannot visit a given interaction bubble more than once. Equivalently, two distinct edges of color $i$ of a $(0i)$-face in a PCGM belong to two distinct interaction bubbles. 
\end{lemma}
\proof Let $\gB$ be a PCGM and assume that there exists a $(0i)$-face $$[\nu_{1}][\nu_{2}][\nu_{3}]\cdots [\nu_{2p-1}][\nu_{2p}]\cdots [\nu_{2q}][\nu_{1}]$$ such that $[\nu_{1}][\nu_{2}]$ and $[\nu_{2p-1}][\nu_{2p}]$ are two edges of color $i$ belonging to the same interaction bubble. Since this bubble is a complete graph, there exists an edge $[\nu_{2p-1}][\nu_{1}]$ of color $j\not = i$. The path $[\nu_{1}][\nu_2]\cdots [\nu_{2p-1}][\nu_{1}]$ is then a cycle of odd length in $\gB_{(0ij)}$. Using Lemma \ref{cyclelemma}, this contradicts the PCGM condition \eqref{leadg}.\qed

Let us denote by $F_{\ell/2}$ the number of $(0i)$-faces of length $\ell$, for any color $i$, where the length of a $(0i)$-face is defined as usual to be the total number of its edges, which is twice the number of edges of color 0. A \emph{self-contraction} is an edge of color 0 attached to two vertices of the same interaction bubble.
\begin{lemma}\label{onefacelemma} A PCGM does not have self-contractions. In particular, $F_{1}=0$.\end{lemma}
\proof Let us assume that the PCGM $\gB$ has a self-contraction and denote by $[\nu_{1}]$ and $[\nu_{2}]$ the two endpoints of the corresponding edge of color 0. Since $[\nu_{1}]$ and $[\nu_{2}]$ belong to the same interaction bubble, there is an edge $[\nu_{1}][\nu_{2}]$ of some color $k$ in this interaction bubble. Choose a pair of colors $(i,j)$ that indexes $[\nu_{1}][\nu_{2}]$ at length two and thus forms a triangle with that edge and a third vertex, say $[\nu_{3}]$. The three-bubble $\gB_{(0ij)}$ then has a cycle of length three, namely $[\nu_{1}][\nu_{3}][\nu_{2}][\nu_{1}]$, which, using Lemma \ref{cyclelemma}, contradicts \eqref{leadg}. The result $F_{1}=0$ immediately follows because the edge of color $0$ in a $(0i)$-face of length two is automatically a self-contraction.

Note that the result can also be obtained immediately from Lemma \ref{revisitlemma}, by considering, for any color $i\not = k$, the $(0i)$-face passing through the vertices $[\nu_{1}]$ and $[\nu_{2}]$.\qed

\vfill\eject

\begin{lemma} \label{lengthface} A PCGM has $F_{2}\geq\frac{1}{2}R(R+1)$.\end{lemma}
\proof By definition, we have
\begin{equation}
\sum_i F_{0i} = \sum_{n\geq 1} F_{n} \, .
\label{Fldef}
\end{equation}
Moreover, each edge of color $0$ belongs to $R$ different $(0i)$-faces, one for each color $i$. The number of edges of color $0$ is the number of propagators $p$ and thus we have
\begin{equation}
\sum_{n\geq 1} n F_{n} = Rp = \frac{R(R+1)}{2} v \, ,
\label{Flrelation}
\end{equation}
where we have used $2p=(R+1)v$ since our Feynman graphs are $(R+1)$-regular. Using  \eqref{leadinggraphs}, \eqref{Fldef} and \eqref{Flrelation} combined with $F_{1}=0$ from Lemma \ref{onefacelemma}, we get
\begin{equation} \label{F2bound}
F_2  = \frac{1}{2}R(R+1) + \frac{1}{4} \sum_{n\geq 3}\Bigl[(R-1)n - 2R-2 \Bigr] F_{n} \, .
\end{equation}
If $R\geq 5$, the second term on the right-hand side is non-negative, which yields the inequality $\smash{F_2 \geq \frac{1}{2} R(R+1)}$. If $R=3$ we further show that $F_{3}=0$ (see also \cite{CarrozzaTanasa} for this case). Indeed, assume that one has a face of length six, say of colors $0$ and $3$. Since there is no self-contraction, this face must visit three distinct interaction bubbles and is thus of the form $[\nu_{1}]^{I}[\nu_{2}]^{I}[\nu_{3}]^{II}[\nu_{4}]^{II}[\nu_{5}]^{III}][\nu_{6}]^{III}[\nu_{1}]^{I}$, where the edges of color 3, namely $[\nu_{1}]^{I}][\nu_{2}]^{I}$, $[\nu_{3}]^{II}[\nu_{4}]^{II}$ and $[\nu_{5}]^{III} [\nu_{6}]^{III}$, belong to three distinct interaction bubbles. Since $R=3$, these three edges are all indexed by the same pair of colors $(1,2)$ at length two, which forms a triangle with each of the edges and third vertices that we call $[\nu_{3/2}]^{I}$, $[\nu_{7/2}]^{II}$ and $[\nu_{11/2}]^{III}$ respectively. The three-colored graph $\gB_{(012)}$ then has a cycle of length nine $[\nu_{1}]^{I}][\nu_{3/2}]^{I}[\nu_{2}]^{I}[\nu_{3}]^{II}[\nu_{7/2}]^{II}[\nu_{4}]^{II}[\nu_{5}]^{III}[\nu_{11/2}]^{III}[\nu_{6}]^{III}[\nu_{1}]^{I}$, contradicting \eqref{leadg} by using Lemma \ref{cyclelemma}.
\qed

As a result, our generalized melons always have several $(0i)$-faces of length four. The following fundamental lemma fixes the structure of these faces.
\begin{lemma}\label{fourfaces} The $(0k)$-faces of length four in a PCGM are of the form $$[\nu_{1}]^{I}[\nu_{2}]^{I}[\nu_{2}]^{II}[\nu_{1}]^{II}[\nu_{1}]^{I}\, ,$$ where $[\nu_{1}]^{I}[\nu_{2}]^{I}$ and $[\nu_{1}]^{II}[\nu_{2}]^{II}$ are two equivalent oriented edges of color $k$ in two distinct interaction bubbles.
\end{lemma}
\proof In the proof, we use the standard labels for the vertices and in particular, vertices $[\nu]^{I}$ and $[\nu]^{II}$ that have the same label $\nu$ are equivalent vertices in two distinct interaction bubbles $I$ and $II$.

There is nothing to prove if $R=3$. We thus assume that $R>3$ and we consider a $(0k)$-face of length four $[\nu_{1}][\nu_{2}][\nu_{3}][\nu_{4}][\nu_{1}]$ in a PCGM, choosing the edges $[\nu_{1}][\nu_{2}]$ and $[\nu_{3}][\nu_{4}]$ to be of color $k$ and thus the edges $[\nu_{2}][\nu_{3}]$ and $[\nu_{4}][\nu_{1}]$ to be of color 0. From Lemma \ref{revisitlemma}, we know that the edges of color $k$ are in two distinct interaction bubbles; we can thus write $[\nu_{1}]=[\nu_{1}]^{I}$, $[\nu_{2}]=[\nu_{2}]^{I}$, $[\nu_{3}]=[\nu_{3}]^{II}$ and $[\nu_{4}]=[\nu_{4}]^{II}$.

Let us first assume that $[\nu_{1}][\nu_{2}]$ and $[\nu_{3}][\nu_{4}]$ are inequivalent unoriented edges. Using Lemma \ref{lengththree}, we can find $(i,j)$ indexing, e.g., $[\nu_{1}][\nu_{2}]$ at length two and $[\nu_{3}][\nu_{4}]$ at length three. Following the $(ij)$-path of length two joining $[\nu_{1}]^{I}$ to $[\nu_{2}]^{I}$, then the edge of color 0 joining $[\nu_{2}]^{I}$ to $[\nu_{3}]^{II}$, then the $(ij)$-path of length three joining $[\nu_{3}]^{II}$ to $[\nu_{4}]^{II}$ and finally the edge of color 0 joining $[\nu_{4}]^{II}$ to $[\nu_{1}]^{I}$, we get a cycle of length seven in $\gB_{(0ij)}$, contradicting planarity by Lemma \ref{cyclelemma} and thus the PCGM condition \eqref{leadg}.

\begin{figure}
\centerline{\includegraphics[width=6in]{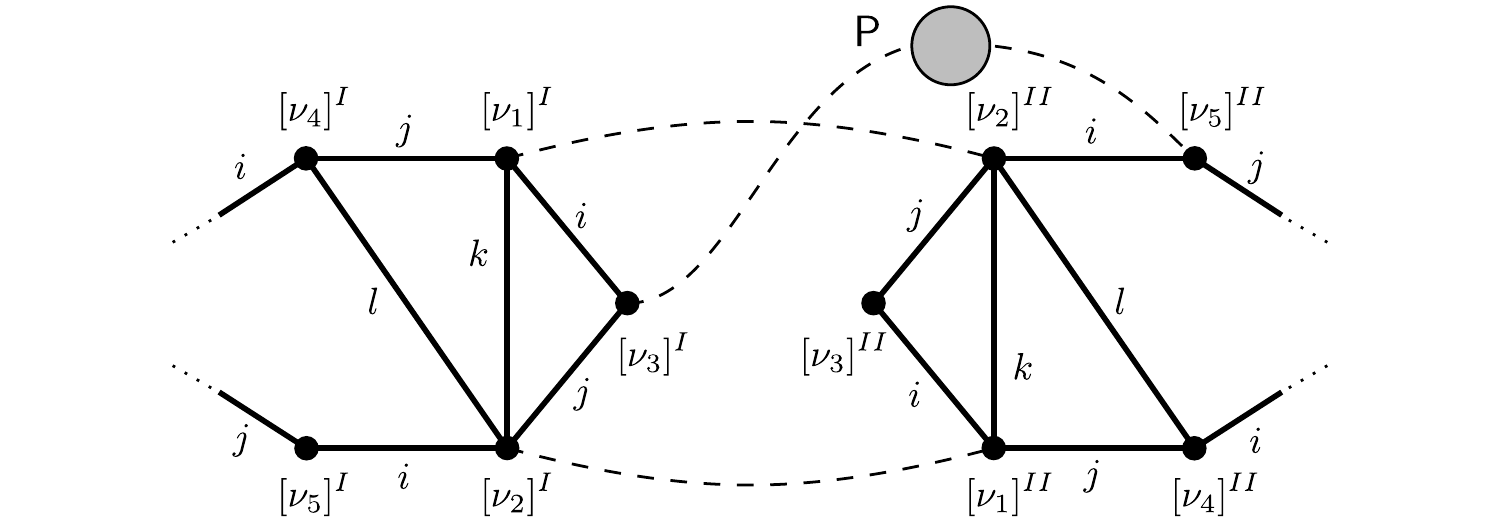}}
\caption{Configuration for which $[\nu_{1}][\nu_{2}]=[\nu_{3}][\nu_{4}]$ and thus $[\nu_{1}]^{I}$ and $[\nu_{2}]^{I}$ are contracted with $[\nu_{2}]^{II}$ and $[\nu_{1}]^{II}$ respectively. Dashed lines represent edges of color 0. Only the relevant parts of the graph are depicted. The $(0i)$-path $\mathsf P$ joining $[\nu_{3}]^{I}$ and $[\nu_{5}]^{II}$ is stylized as a grey disk attached to two edges of color 0. This path is represented in more details in Fig.\ \ref{figproofb}.}\label{figproofa}
\end{figure}
\begin{figure}[ht]
\centerline{\includegraphics[width=6in]{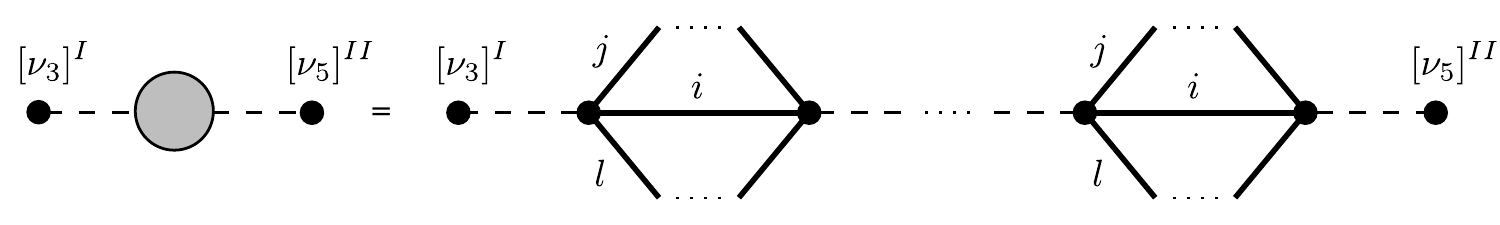}}
\caption{$(0i)$-path $\mathsf P$ joining $[\nu_{3}]^{I}$ to $[\nu_{5}]^{II}$, together with the $(jl)$-faces in the various distinct interaction bubbles visited by the path.}\label{figproofb}
\end{figure}

There remains two possibilities: $[\nu_{1}][\nu_{2}]=[\nu_{4}][\nu_{3}]$, which is what we want to prove, or $[\nu_{1}][\nu_{2}]=[\nu_{3}][\nu_{4}]$. Let us assume that the second possibility is realized, which means that $[\nu_{1}]^{I}$ and $[\nu_{2}]^{I}$ are contracted with $[\nu_{2}]^{II}$ and $[\nu_{1}]^{II}$ respectively. The resulting configuration is depicted in Fig.\ \ref{figproofa}. The important features are as follows. We have indexed the edge $[\nu_{1}][\nu_{2}]$ at length two by the pair of colors $(i,j)$, with associated triangles $[\nu_{1}][\nu_{2}][\nu_{3}]$ in the two interaction bubbles. We have depicted part of the $(ij)$-face in both interaction bubbles, introducing in particular the vertices $[\nu_{4}]$ and $[\nu_{5}]$ and the edge $[\nu_{2}][\nu_{4}]$. The color of this edge is denoted by $l$. Note that $i$, $j$, $k$ and $l$ must be four distinct colors. We have also explicitly depicted the $(0i)$-face that contains the edges $[\nu_{1}]^{I}[\nu_{3}]^{I}$ and $[\nu_{2}]^{II}[\nu_{5}]^{II}$. From Lemma \ref{revisitlemma}, we know that the $(0i)$-path joining $[\nu_{3}]^{I}$ to $[\nu_{5}]^{II}$ along this face, which we call $\mathsf P$, visits distinct interaction bubbles at each intermediate edge of color $i$ (so, for example, the edge $[\nu_{2}]^{I}[\nu_{1}]^{II}$ of color 0 cannot belong to this path). We have represented in more details the path $\mathsf P$ in Fig.\ \ref{figproofb}, indicating as well the $(jl)$-faces in each intermediate interaction bubble visited by the path. 

\begin{figure}
\centerline{\includegraphics[width=6in]{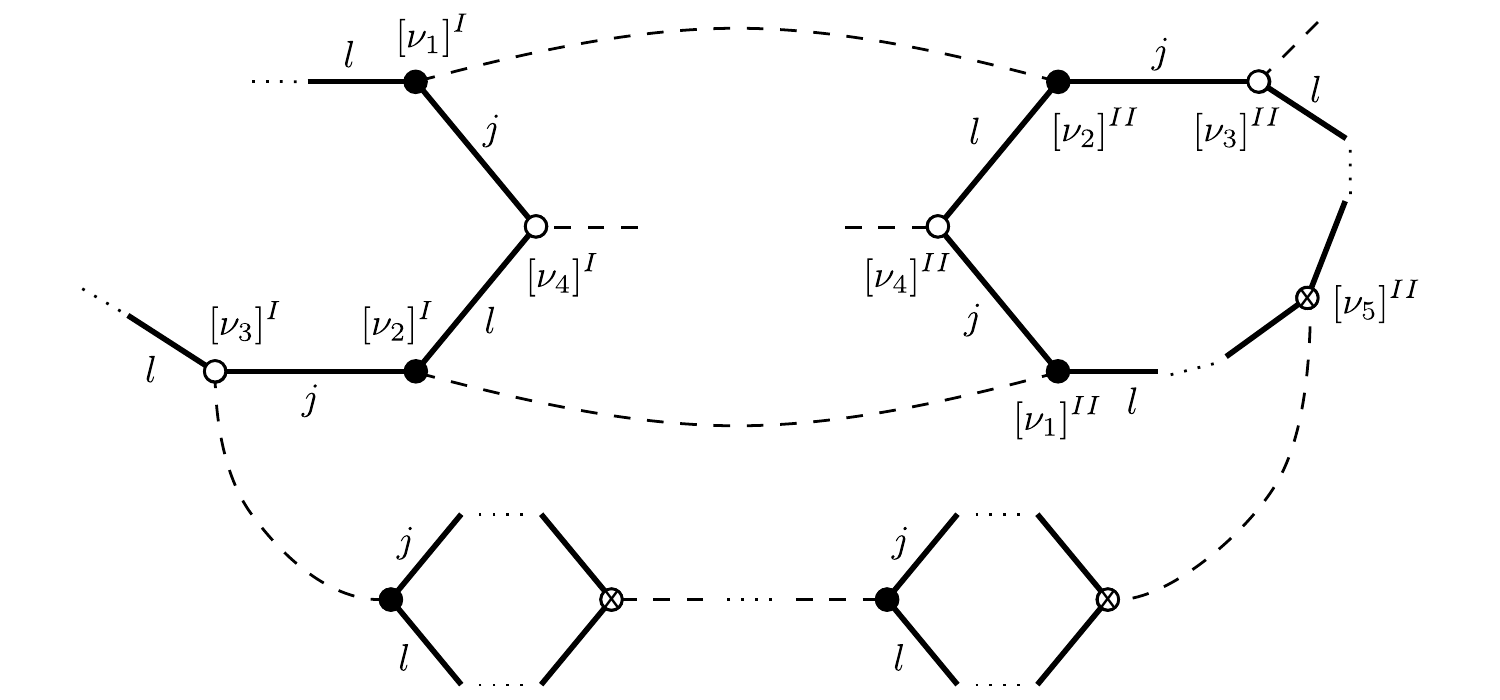}}
\caption{Three-bubble $\gB_{(0jl)}$, deduced from the graphs depicted in Fig.\ \ref{figproofa} and \ref{figproofb}, drawn with a convenient choice of embedding. The type (filled or unfilled) of the vertices indicated with a cross is a priori unknown.}\label{figproofc}
\end{figure}

\begin{figure}
\centerline{\includegraphics[width=6in]{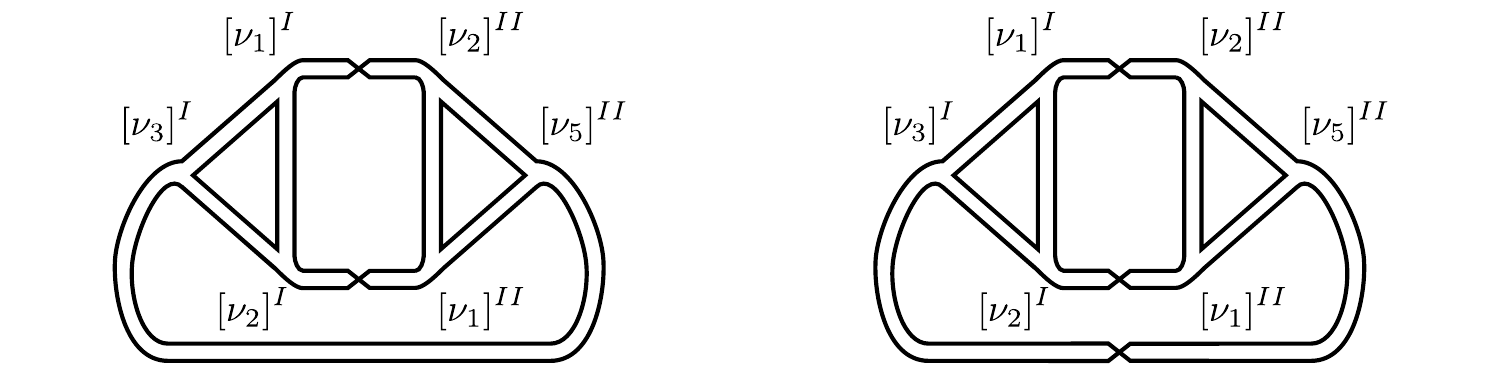}}
\caption{The two possible ribbon graphs obtained from the three-bubble $\gB_{(0jl)}$ depicted in Fig.\ \ref{figproofc} after performing the edge and vertex deletion operations described in the main text. Both have genus one, contradicting planarity.}\label{figproofd}
\end{figure}

Consider now the three-bubble $\gB_{(0jl)}$. A convenient representation, obtained starting from the graphs in Fig.\ \ref{figproofa} and \ref{figproofb}, is given in Fig.\ \ref{figproofc}. Without loss of generality, the embedding is chosen such that the edges of color 0 attached to the polygonal $(jl)$-faces always point outwards. This implies that the ribbons $[\nu_{1}]^{I}[\nu_{2}]^{II}$ and $[\nu_{2}]^{I}[\nu_{1}]^{II}$ are twisted. Because we are in a PCGM, we know that $\gB_{(0jl)}$ must be planar. Let us then use the edge and vertex deletion operations described in Lemma \ref{oplemma} in the following way: we first delete all the edges of color 0, except for $[\nu_{1}]^{I}[\nu_{2}]^{II}$, $[\nu_{2}]^{I}[\nu_{1}]^{II}$ and the edges in the path $\mathsf P$ that join the distinct interaction bubbles within this path; we then delete pieces of each $(jl)$-faces in $\mathsf P$ (the upper or lower parts in each interaction bubble when the path is depicted as in Fig.\ \ref{figproofc}) so that the path is reduced to a succession of ribbons attached to vertices of valency two; finally, we delete all the vertices of valency two. Taking into account the fact that the type of some vertices is a priori unknown in the embedding, these operations can produce one of the two ribbon graphs depicted in Fig.\ \ref{figproofd}. From Lemma \ref{oplemma}, at least one of these graphs must be planar. However, it is easy to check that they both have genus one. Our initial hypothesis, that $[\nu_{1}][\nu_{2}]=[\nu_{3}][\nu_{4}]$, is thus impossible. This concludes the proof.\qed

\subsubsection{The PCGM with two bubbles} \label{subsectiontwovertices}

A PCGM with $v=1$ cannot exist, because it would necessarily have self-contractions. The simplest PCGMs thus have $v=2$, which is the case considered in the present subsection. We use the standard vertex labeling, with vertices $[C]^{I}$, $[k]^{I}$ and $[C]^{II}$, $[k]^{II}$, $1\leq k\leq R$, for the interaction bubbles number one and two respectively.

\begin{lemma}\label{twoverticeslemma} There exists a unique PCGM with $v=2$, called the elementary generalized melon, corresponding to the symmetric configuration (i.e.\ elementary mirror melon, see Sec.\ \ref{MirrorSec}) where edges of color 0 join $[C]^{I}$ to $[C]^{II}$ and $[k]^{I}$ to $[k]^{II}$ for all $1\leq k\leq R$. In the case $R=3$, this elementary generalized melon is obtained from four distinct Wick contractions between the two prime-complete interaction bubbles, whereas for $R>3$ it is obtained from a unique Wick contraction.
\end{lemma}
\proof Lemmas \ref{revisitlemma}, \ref{onefacelemma} and \ref{fourfaces} immediately fix the PCGM with $v=2$ to the symmetric configuration. Note that this is of course consistent with \eqref{leadinggraphs} and the inequality $F_{2}\geq \frac{1}{2}R(R+1)$, which predict that a PCGM with $v=2$ has precisely $F_{2}=\frac{1}{2}R(R+1)$ and $F_{n}=0$ if $n\not =2$.

When $R>3$, since all the vertices of the prime-complete bubble are inequivalent, there is clearly a unique Wick-contraction that yields this symmetric configuration. When $R=3$, since all the vertices are equivalent, one can start by Wick-contracting any given vertex of the first bubble with any vertex of the second bubble, yielding four possibilities. It is straightforward to check that, after this initial choice is made, the other contractions are automatically fixed by the requirement that all the $(0i)$-faces have length four. Because of the special symmetry properties of $\mathcal K_{4}$, the four graphs we get in this way are actually four copies of the same elementary generalized melon.\qed

\begin{figure}
\centerline{\includegraphics[width=6in]{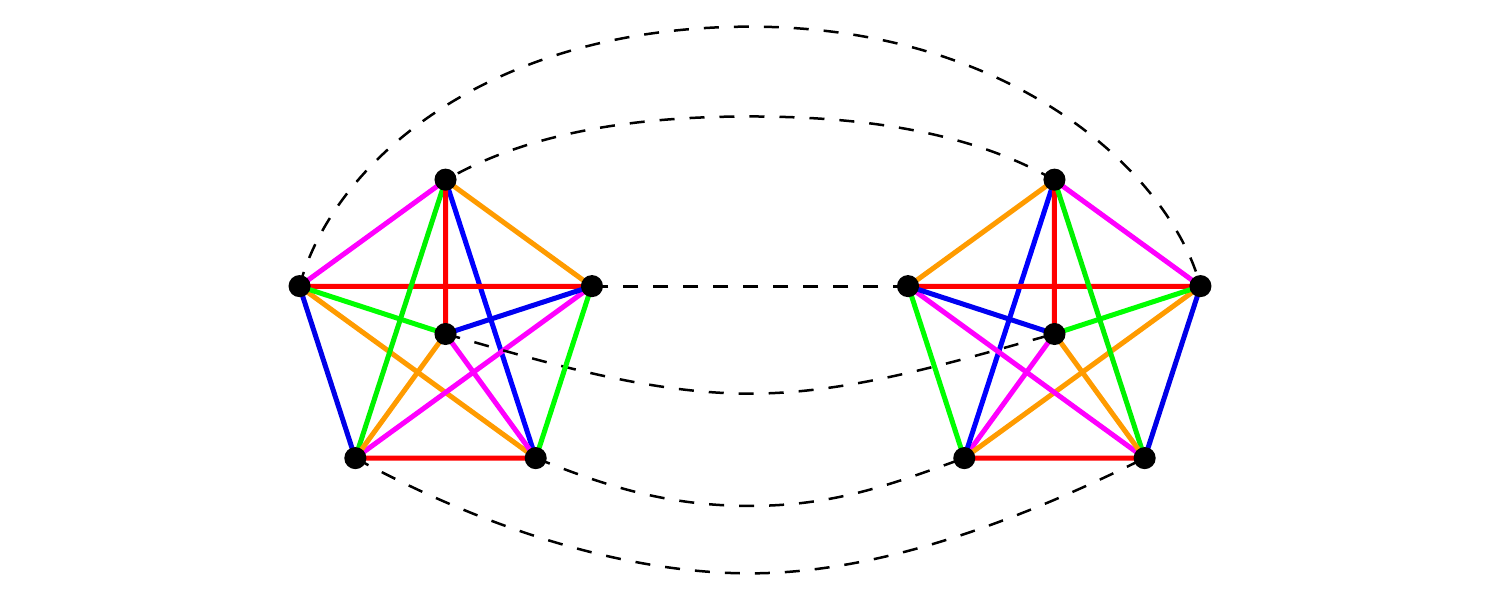}}
\caption{Unique PCGM with two interaction bubbles, also called the elementary generalized melon, in the case $R=5$. The dashed lines are the edges of color 0.}
\label{figureB}
\end{figure}

The resulting elementary generalized melon is depicted in Fig.\ \ref{figureB} in the case $R=5$.

\begin{figure}[ht]
\centerline{\includegraphics[width=6in]{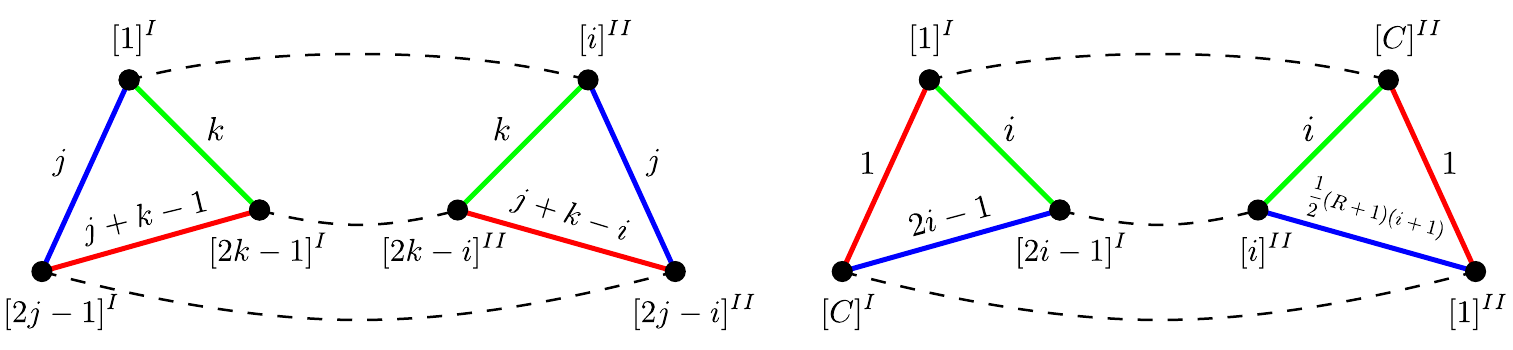}}
\caption{\label{proofLemFig1} Configurations used in the elementary proof of Lemma \ref{twoverticeslemma}. Consistency of the left picture requires $i=1$, whereas the right picture is impossible.}
\end{figure}

For completeness, let us also mention a completely elementary proof of Lemma \ref{twoverticeslemma} that does not use the non-trivial Lemma \ref{fourfaces} but only the fact that all the $(0i)$-faces must be of length four. It goes as follows (see Fig.\ \ref{proofLemFig1}).

Let us first assume that $[1]^{I}$ is Wick-contracted with $[i]^{II}$. We then consider two distinct edges $[1]^{I}[2j-1]^{I}$ and $[1]^{I}[2k-1]^{I}$, which have respectively colors $j$ and $k$ both different from $i$ and 1. This is always possible because $R>3$. The $(0j)$-face containing $[1]^{I}[2j-1]^{I}$ is then of length four if and only if $[2j-1]^{I}$ is contracted with the vertex $[2j-i]^{II}$, so that $[1]^{I}[2j-1]^{I}$ and $[i]^{II}[2j-i]^{II}$ have the same color $j$ (note that $[2j-1]^{I}$ cannot be contracted with the vertex $[C]^{II}$, because by choice $j\not = i$). Similarly, $[2k-1]^{I}$ must be contracted with $[2k-i]^{II}$. The face of colors 0 and $j+k-1$ containing the edge $[2j-1]^{I}[2k-1]^{I}$ is then of length four if and only if the edge $[2j-i]^{II}[2k-i]^{II}$ is of color $j+k-1$, which yields $i\equiv 1$.

Let us second assume that $[1]^{I}$ is Wick-contracted with $[C]^{II}$. The face containing $[C]^{I}[1]^{I}$ is of length four if and only if $[C]^{I}$ is contracted with $[1]^{II}$. Let us also consider the $(0i)$-face, $i\not = 1$, containing the edge $[1]^{I}[2i-1]^{I}$. It is of length four if and only if $[2i-1]^{I}$ is Wick-contracted with $[i]^{II}$. But then, the face of colors $0$ and $2i-1$ containing the edge $[C]^{I}[2i-1]^{I}$ is of length four if and only if the color of the edge $[1]^{II}[i]^{II}$ is $2i-1$, which yields $3i\equiv 3$. For $R>3$, this implies $i\equiv 1$, which is impossible.

We thus conclude that, for $R>3$, $[1]^{I}$ must be contracted with $[1]^{II}$. Exactly the same reasoning shows that $[k]^{I}$ must be contracted with $[k]^{II}$ for all $1\leq k\leq R$. The two center vertices $[C]^{I}$ and $[C]^{II}$ are then also automatically contracted.\qed

In the following, it will also be useful to consider ``elementary generalized two-point melons,'' which are obtained from the elementary generalized melon by cutting open an edge of color 0. Note that, since such an edge belongs to exactly one face of colors 0 and $i$, for a given $i$, and the original elementary generalized melon contains $\frac{1}{2}(R+1)$ faces of colors 0 and $i$, an elementary generalized two-point melon itself contains $\frac{1}{2}(R+1) - 1 = \frac{1}{2}(R-1)$ faces of colors 0 and $i$, for any given $i$.

\subsubsection{The most general PCGMs}
\begin{figure}
\centerline{\includegraphics[width=6in]{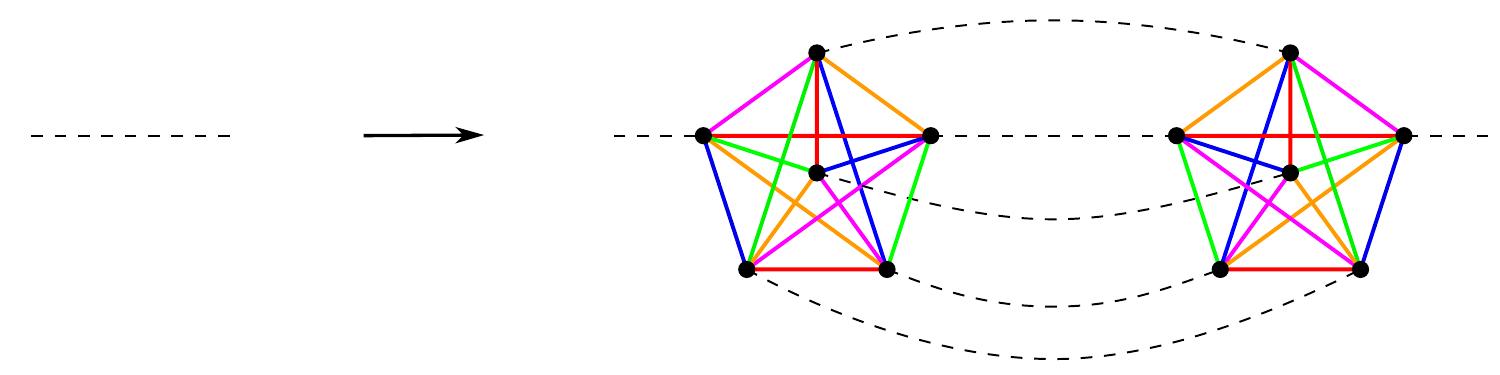}}
\caption{Example of a generalized melonic insertion, in the case $R=5$. This operation does not change the index of a Feynman graph, see the discussion in Sec.\ \ref{gmSec}.}
\label{figureC}
\end{figure}

By combining the results of Lemmas \ref{twoverticeslemma} and the generalized melonic moves depicted in Fig.\ \ref{figureC} (see Sec.\ \ref{gmSec}), we can build an infinite family of PCGMs, starting from the elementary generalized melon and using an arbitrary number of melonic insertions. We are now going to prove that the most general PCGMs can be obtained in this way.

A prime-complete generalized two-point melon, or PCG2M for short, is defined to be the graph obtained from a PCGM by cutting open any edge of color 0. The trivial PCG2M is simply a single edge of color 0.

\begin{figure}
\centerline{\includegraphics[width=6in]{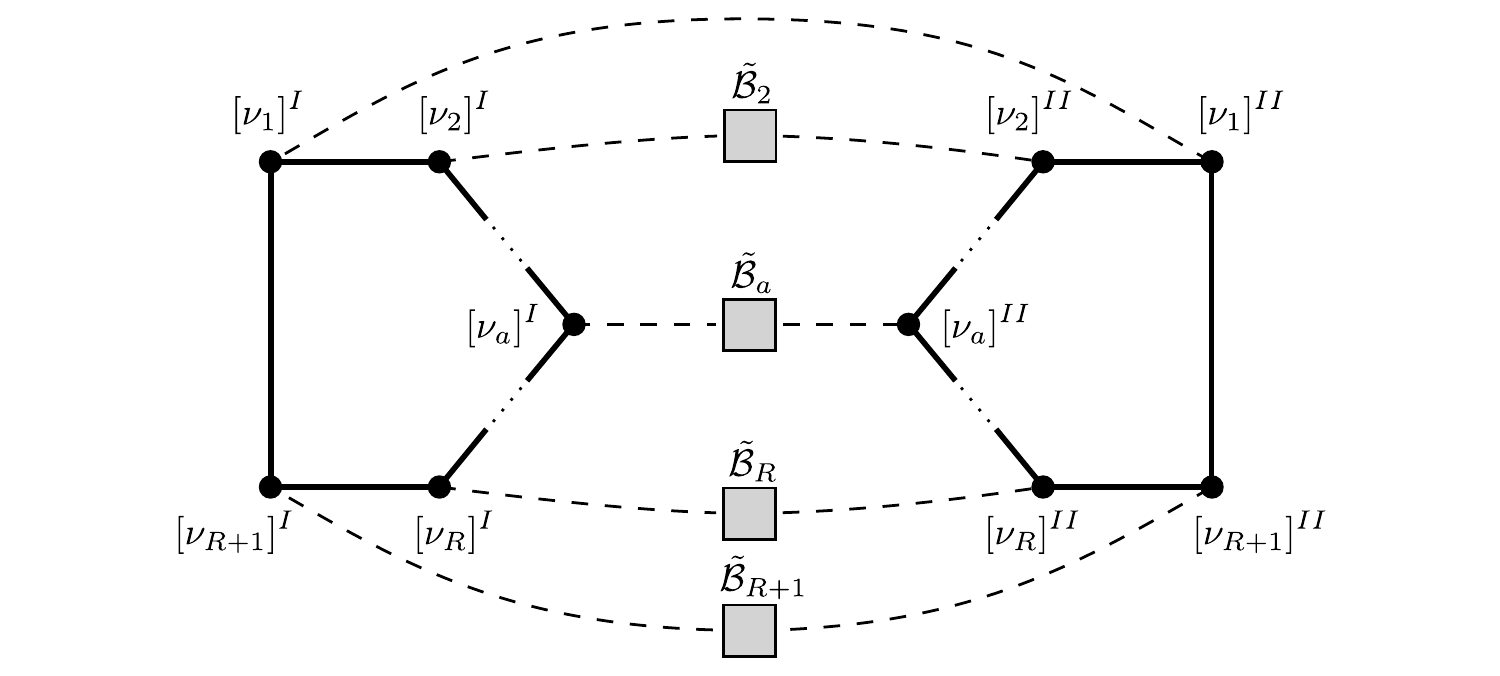}}
\caption{General structure of a PCGM in the form of $R+1$ PCG2Ms $\tilde\gB_{a}$, $1\leq a\leq R+1$, each attached to equivalent vertices $[\nu_{a}]^{I}$ and $[\nu_{a}]^{II}$ in two distinct interaction bubbles. The bubbles $\gK_{R+1}$ are depicted sketchily in an arbitrary $(ij)$-polygonal representation for which only the boundary of the polygon is drawn. The PCG2Ms are stylized as light grey squares attached to two edges of color 0. At least one of the $\tilde\gB_{a}$, $2\leq a\leq R+1$, is trivial.}
\label{fig2M}
\end{figure}
\begin{lemma}\label{thlemma1} Any PCGM can be represented in the form of $R+1$ PCG2Ms, with at least two of them being trivial, each attached to equivalent vertices in two distinct interaction bubbles $\gK_{R+1}$, see Fig.\ \ref{fig2M}.
\end{lemma}

\begin{figure}
\centerline{\includegraphics[width=6in]{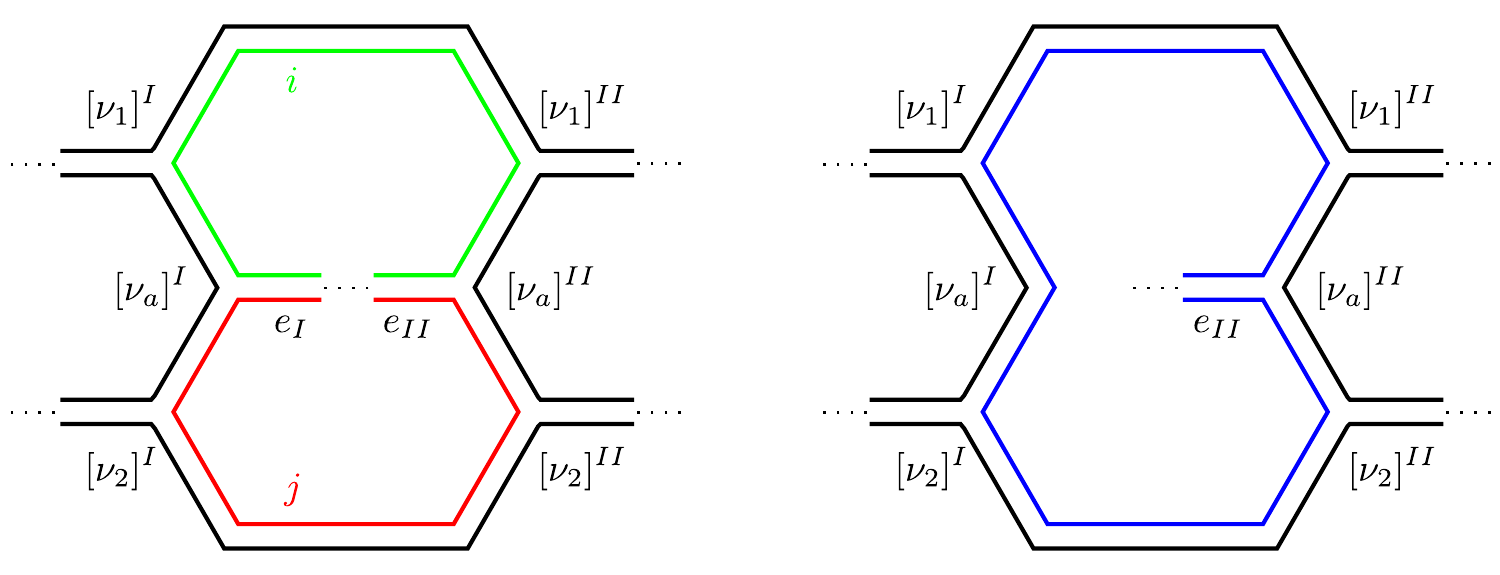}}
\caption{Ribbon graph for the planar three-bubble $\gB_{(0ij)}$ used in the proof of Lemma \ref{thlemma1} (left inset) and connected planar ribbon graph obtained after deletion of the regular edge $e_{I}$ (right inset). The green and red faces merge into a unique blue face and the edge $e_{II}$ becomes singular.}
\label{fig3M}
\end{figure}

\proof Let $\gB$ be a PCGM. We start by using a $(0k)$-face of length four, whose existence is ensured by Lemma \ref{lengthface}. The structure of this face is given by Lemma \ref{fourfaces}. This provides our two interaction bubbles $I$ and $II$. We then pick any vertex $[\nu_{a}]$ in $\gK_{R+1}$ different from $[\nu_{1}]$ and $[\nu_{2}]$. The edges $[\nu_{1}][\nu_{a}]$ and $[\nu_{a}][\nu_{2}]$ have colors $i$ and $j$ respectively. We call $e_{I}$ and $e_{II}$ the edges of color 0 attached to the equivalent vertices $[\nu_{a}]^{I}$ and $[\nu_{a}]^{II}$. As usual, the PCGM condition \eqref{leadg} implies that the three-bubble $\gB_{(0ij)}$ must be planar. The associated ribbon graph, in a convenient embedding, is depicted on the left of Fig.\ \ref{fig3M}. We have outlined in green and red the $(0i)$- and $(0j)$-faces containing $[\nu_{1}][\nu_{a}]$ and $[\nu_{a}][\nu_{2}]$ respectively.

\begin{figure}
\centerline{\includegraphics[width=6in]{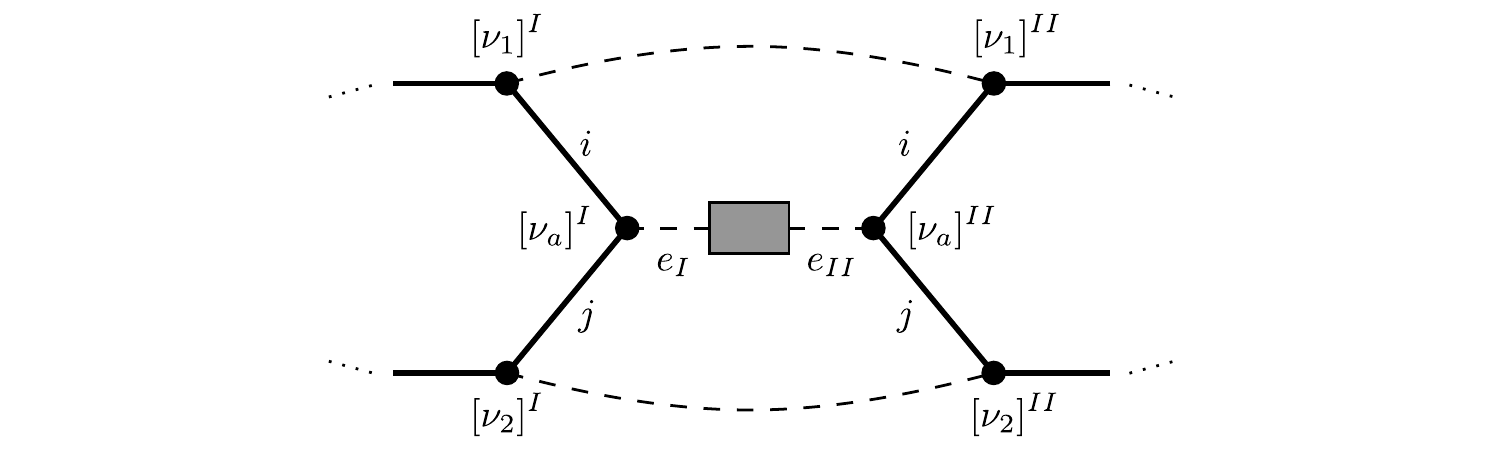}}
\caption{Structure of $\gB_{(0ij)}$, as implied by its two-particle reducibility with respect to the edge $e_{I}$ and $e_{II}$. The graph $\gB$ itself must have a similar structure.}
\label{fig4M}
\end{figure}

Let us now delete the regular edge $e_{I}$. This yields the ribbon graph depicted on the right of Fig.\ \ref{fig3M}. From Lemma \ref{oplemma}, we know that this graph must be connected and planar. In this new graph, the edge $e_{II}$ is singular, because the original green and red faces have merged together. Using again Lemma \ref{oplemma}, we conclude that the deletion of the edge $e_{II}$ produces two connected planar components. In other words, $\gB_{(0ij)}$ is two-particle reducible with respect to the edges $e_{I}$ and $e_{II}$. The structure of $\gB_{(0ij)}$ must then be as illustrated in Fig.\ \ref{fig4M}. The dark grey rectangular region in this figure represents one of the connected planar components obtained after deleting $e_{I}$ and $e_{II}$.

Because the interaction bubbles $\gK_{R+1}$ are MST, it is obvious that the connectivity properties of $\gB$ and $\gB_{(0ij)}$ are the same. In particular, one can find a path that joins two vertices in $\gB$ and that does not contain the edges $e_{I}$ and $e_{II}$ if and only if the same is true in $\gB_{(0ij)}$. Therefore, after the deletion of the edges $e_{I}$ and $e_{II}$, the graph $\gB$ itself splits into two connected components. A picture similar to the one for $\gB_{(0ij)}$ on Fig.\ \ref{fig4M} is thus valid for $\gB$ as well. Moreover, since the equivalent vertices $[\nu_{a}]^{I}$ and $[\nu_{a}]^{II}$ were chosen arbitrarily, we can repeat the argument for all the pairs of equivalent vertices in the interaction bubbles $I$ and $II$. Eventually, we obtain the picture of Fig.\ \ref{fig2M}. We also know that one of the $\tilde\gB_{a}$, for some $2\leq a\leq R+1$, is trivial.\footnote{In Fig.\ \ref{fig2M}, this trivial $\tilde\gB_{a}$ is not necessarily $\tilde\gB_{2}$, because the polygon boundaries used in this figure do not necessarily correspond to the $(ij)$-faces used in the proof.}

\begin{figure}
\centerline{\includegraphics[width=6in]{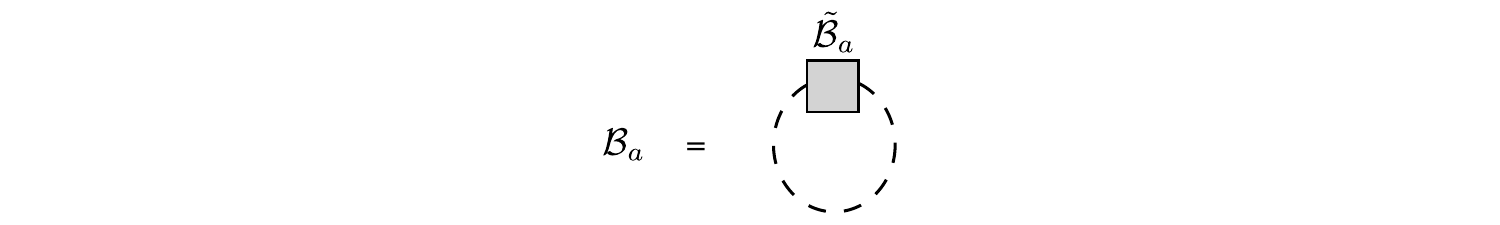}}
\caption{Bubbles $\gB_{a}$ constructed from the two-point graphs $\tilde\gB_{a}$.}
\label{fig5M}
\end{figure}

It remains to prove that the $\tilde\gB_{a}$ are all PCG2Ms. This is equivalent to the fact that the bubbles $\gB_{a}$ depicted in Fig.\ \ref{fig5M} are PCGMs, which is itself equivalent to the planarity of the three-colored graphs $(\gB_{a})_{(0ij)}$ for all pairs of colors $(i,j)$. Then, let us pick two colors $i$ and $j$ and consider $\smash{\gB_{(0ij)}}$. The latter corresponds to a planar graph that looks like the one depicted in Fig.\ \ref{fig2M}, the two polygon boundaries being the two $(ij)$-faces and the $\tilde\gB_{a}$ being replaced by the graphs $\smash{(\tilde\gB_{a})_{(0ij)}}$ obtained from $\tilde\gB_{a}$ by keeping the edges of colors 0, $i$ and $j$ only. If we delete all the edges of color 0 except $[\nu_{1}]^{I}[\nu_{1}]^{II}$ and the two attached to $[\nu_{a}]^{I}$ and $[\nu_{a}]^{II}$, then all the vertices of valency two and then two more edges, one joining $[\nu_{1}]^{I}$ to $[\nu_{a}]^{I}$ and the other $[\nu_{1}]^{II}$ to $[\nu_{a}]^{II}$, we get precisely the graph $(\gB_{a})_{(0ij)}$. Since we started from the planar graph $\gB_{(0ij)}$, Lemma \ref{oplemma} implies that $(\gB_{a})_{(0ij)}$ must be planar too and we conclude.\qed

We can now state and easily prove the main result of the present section.

\begin{theorem} \label{theorem1}
The most general PCGMs, which are the leading order vacuum graphs of the model defined by the action \eqref{actionGenCT} with \eqref{uncoloredint} and $R$ prime, are obtained by performing an arbitrary number of generalized melonic insertions starting from the elementary generalized melon.\end{theorem}
\proof Let $v$ be the total number of interaction bubbles in a PCGM. From Lemmas \ref{onefacelemma} and \ref{twoverticeslemma}, we know that the theorem is true for $v\leq 2$. We then proceed recursively. Assume that it is true for all $v<v_{0}$ and consider a PCGM $\gB$ having $v_{0}$ interaction bubbles. We use Lemma \ref{thlemma1} to put it in the form of Fig.\ \ref{fig2M}. The PCGMs $\gB_{a}$, built from the $\tilde\gB_{a}$ as shown in Fig.\ \ref{fig5M}, have  at most $v_{0}-2$ interaction bubbles. The theorem follows by using the recursion hypothesis on these PCGMs.\qed

In conclusion, the PCGMs coincide exactly with the mirror melons defined in Sec.\ \ref{MirrorSec}.

\subsubsection{\label{RnotprimeSec}Remark on the case of \texorpdfstring{$R$}{R} not prime}

We now return to the model defined by \eqref{actionGenCT} but in the case of $R$ not prime, so that the complete
interaction bubble $\gK_{R+1}$ is not MST. For any standard coloring, we can build the mirror melons of Sec.\ \ref{MirrorSec} associated to the complete bubble.
It is easy to compute their number of faces. By induction, a mirror melon with $v$ vertices has exactly 
$R + v \frac{R(R-1)}{4}$ faces of color 0 and $i$, for any color $i$. Hence, by \eqref{degreeintnotprime}, all mirror melons have strictly positive index and are not generalized melons. In particular, the scaling of \eqref{actionGenCT} for $R$ not prime is strictly enhanced compared to the one of \eqref{newcouplings}. It is the right one for mirror melons to scale as $N^R$, independently of their number of vertices. However we do not know if the leading sector is made solely of the mirror melons in this case, or even if the large $N$ limit makes sense.

\subsection{Application: quantum models and SYK physics} 

The prime-complete interaction can be used to build interesting quantum mechanical (and field theoretic) matrix-tensor theories that can model quantum black holes. We are going to briefly discuss two Hamiltonians, one based on Majorana fermions and the other on Dirac fermions.

\subsubsection{\label{MajoranaSec}Prime-complete Majorana fermion model}

We consider real fermionic matrix-tensor operators $$\psi_{ab\mu_{1}\cdots\mu_{r}}=(\psi_{ab\mu_{1}\cdots\mu_{r}})^{\dagger}\, ,\quad 1\leq a,b\leq N\, ,\ 1\leq\mu_{i}\leq D\, ,$$ satisfying the quantization conditions
\be\label{qc1}\bigl\{\psi_{ab\mu_{1}\cdots\mu_{r}},\psi_{cd\nu_{1}\cdots\nu_{r}}\bigr\} = \frac{1}{ND^{r}} \delta_{ac}\delta_{bd}\delta_{\mu_{1}\nu_{1}}\cdots
\delta_{\mu_{r}\nu_{r}}\, .\ee
The $\text{O}(N)^{2}\times\text{O}(D)^{r}$ symmetric Hamiltonian is
\be\label{Hdef} H=-\frac{1}{2} i^{\frac{1}{2}(r+2)(r+3)}ND^{r+\frac{1}{4}r(r+1)}\la\tr\Bigl(\psi_{[C]}\psi^{T}_{[2]}\prod_{p=1}^{\frac{r+1}{2}}\psi_{[2-2p]}\psi^{T}_{[2+2p]}\Bigr) + \text{H. c.}
\ee
We use here the matrix notation associated with the colors $1$ and $2$, the trace in the Hamiltonian being associated with the $(12)$-face of the prime-complete interaction bubble $\gK_{r+3}$. We have indexed the variables according to which vertex they are associated to in $\gK_{r+3}$. The appropriate contractions of the $\text{O}(D)$ indices are assumed. For example,
\begin{align}\label{H1ex} H &= ND^{\frac{3}{2}}\la_{1}\tr\psi_{\mu}\psi_{\nu}^{T}\psi_{\mu}\psi_{\nu}^{T}\quad\text{for $r=1$}\, ,\\\label{H3ex}
H&= \frac{i}{2}ND^{6}\la\tr\psi_{\alpha\mu\theta}\psi^{T}_{\beta\nu\phi}\psi_{\gamma\rho\theta}\psi^{T}_{\beta\mu\xi}\psi_{\alpha\rho\phi}\psi^{T}_{\gamma\nu\xi} + \text{H. c.}\quad\text{for $r=3$}\, .
\end{align}
Note that for $r=1$, only $\la_{1}=\re\la$ contributes, whereas for $r\geq 3$ it is essential to add the Hermitian conjugate term to get a unitary theory. This is related to the fact that the vertices are all inequivalent in $\gK_{r+3}$ for $r\geq 3$, as discussed in Sec.\ \ref{disbubbleSec}. However, the difference between the two terms in the Hamiltonian is a subleading effect at large $N$ and large $D$. The factor of $i$ in \eqref{H3ex} has been chosen so that only $\la_{1}=\re\la$ contributes at leading order, for any $r$.

The basic quantity in the model is the Euclidean two-point function at finite temperature
\be\label{Gdef1} G(t) = \frac{1}{N}\bigl\langle\tr\psi_{\mu_{1}\cdots\mu_{r}}(t)\psi^{T}_{\mu_{1}\cdots\mu_{r}}\bigr\rangle_{\beta}\, .\ee
It is a real and odd function of the Euclidean time $t$ satisfying $G(t+\beta)=-G(t)$. From our classification theorem \ref{theorem1}, it is straightforward to write down the Schwinger-Dyson equation that determines $G$. We introduce as usual the Fourier transform
\be\label{GFourier} G(t) = \frac{1}{\beta}\sum_{k\in\mathbb Z+\frac{1}{2}}G_{k}\,e^{-\frac{2i\pi k t}{\beta}}\ee
and the self-energy $\Sigma$,
\be\label{Sigmadef} \frac{1}{G_{k}}=-\frac{2i\pi k}{\beta} + \Sigma_{k}\, ,\quad \Sigma(t) = \frac{1}{\beta}\sum_{k\in\mathbb Z+\frac{1}{2}}\Sigma_{k}\,e^{-\frac{2i\pi k t}{\beta}}\, .\ee
The Schwinger-Dyson equation then reads
\be\label{SDMajorana} \Sigma (t) = 
\begin{cases} -16 \la_{1}^{2}G(t)^{3} & \text{if $r=1$,}\\
-(r+3)\la_{1}^{2}G(t)^{r+2} & \text{if $r\geq 3$ and $r+2$ is prime.}
\end{cases}\ee
The fact that the combinatorial factor 16 when $r=1$ does not generalize to $(r+3)^{2}$ for larger values of $r$ is again an effect of the lack of symmetry between the vertices in the bubble for $r\geq 3$.

Several comments are in order.

\noindent i) We have obtained a matrix-tensor version of the generalized version of the SYK model considered in \cite{maldastan} and we can in particular study the limit $q=r+3\rightarrow\infty$. Note, however, that the result holds in principle only when $r+2=q-1$ is a prime number. As mentioned in Sec.\ \ref{RnotprimeSec}, we do not know the form of the leading graphs and thus we cannot write down the Schwinger-Dyson equation if $r+2$ is not prime. It is a logical possibility that \eqref{SDMajorana} is still correct, but our proof would have to be generalized non-trivially, since the relation with the index is then lost. 

\noindent ii) In the case of the SYK model, a straightforward and purely algebraic derivation of the leading order Schwinger-Dyson equation from standard manipulations of the path integral with an auxiliary field is possible, without any study of the Feynman diagrams themselves \cite{maldastan,jevicki}. An analogue of such a derivation is not known for our matrix-tensor models and most probably does not exist. The reason is that when there is a formulation in terms of an auxiliary field, the large $N$ expansion \emph{at all orders and for all the correlation functions} can be obtained straightforwardly from the loop expansion generated by the effective action for the auxiliary field. In cases like \eqref{Hdef} the study of subleading orders is typically much more difficult and cannot be reduced to a simple loopwise expansion. For a discussion of subleading orders in related models, see e.g.\ \cite{gurauwitten,Bonzom:2017pqs}.

\noindent iii) In spite of the lack of an auxiliary field formulation, all the thermodynamical functions (free energy, etc.) in our model can be expressed at leading order in terms of the two-point function \eqref{Gdef1} only, see \cite{AFS}.

\noindent iv) The fact that a tensor model mimicking the $q$-fold random SYK interaction can be built has been mentioned recently in the literature \cite{Narayan:2017qtw,qfolf}. These papers seem to have assumed that the analysis of the simplest case in \cite{CarrozzaTanasa} could be immediately generalized. As we have seen, this is incorrect. Still, as we have said, it might be that the Schwinger-Dyson equation \eqref{SDMajorana} remains valid (plausibly modulo appropriate combinatorial factors) beyond the case $R$ prime which is fully solved in the present paper. For $R$ prime, the particular combinatorial factors in \eqref{SDMajorana} are related to the fact that the vertices of the prime-complete bubble are distinguishable when $r\geq 3$. These factors are not correctly written down in \cite{Narayan:2017qtw,qfolf}. However, this seems to be irrelevant for the bulk of the results in these references.

%
%
%
%
%
%

%
\subsubsection{\label{DiracSec}Prime-complete Dirac fermion model}

The SYK model has an interesting complex version studied in particular in \cite{Sachdev}; it is natural to also consider a complex version of the matrix-tensor model \eqref{Hdef}. An important physical motivation to do so is to be able to add a non-trivial mass term. The resulting phase diagram when $r=1$ has been shown recently to display many surprising features, the most notable being the existence of a non-trivial critical point \cite{AFS}. The generalization of the model to any prime $R=r+2$ opens the way to a detailed analytical study of this critical point in the large $r$ limit \cite{AFS}.

The model is built from complex matrix-tensor operators satisfying the quantization conditions
\be\label{qc2}\bigl\{\psi^{a}_{\mu_{1}\cdots\mu_{r}\, b},(\psi^{\dagger}_{\nu_{1}\cdots\nu_{r}})^{c}_{\ d}\bigr\}= \frac{1}{ND^{r}} \delta^{a}_{d}\delta^{c}_{b}\delta_{\mu_{1}\nu_{1}}\cdots
\delta_{\mu_{r}\nu_{r}}\, .\ee
We use the convention $$(\psi^{a}_{\mu_{1}\cdots\mu_{r}\, b})^{\dagger}=(\psi^{\dagger}_{\mu_{1}\cdots\mu_{r}})^{b}_{\ a}\, ,\quad 1\leq a,b\leq N\, ,\ 1\leq\mu_{i}\leq D.$$ The $\text{U}(N)^{2}\times\text{O}(D)^{r}$ symmetric Hamiltonian is
\be\label{Hdef2} H= ND^{r}\tr\biggl[ m \psi^{\dagger}_{\mu_{1}\cdots\mu_{r}}\psi_{\mu_{1}\cdots\mu_{r}} + D^{\frac{1}{4}r(r+1)}\Bigl(\la
\psi_{[C]}\psi^{\dagger}_{[2]}\prod_{p=1}^{\frac{r+1}{2}}\psi_{[2-2p]}\psi^{\dagger}_{[2+2p]}
+ \text{H. c.}\Bigr)\biggr]\, ,\ee
with notations similar to what we used in \eqref{Hdef}. When $r=1$, we get the model studied in \cite{AFS}.\footnote{The coupling $\la$ in \eqref{Hdef2} is $\frac{1}{4}$ the coupling $\la$ in \cite{AFS}.} When $r\geq 3$, the addition of the Hermitian conjugate term is essential for unitarity. Unlike in \eqref{Hdef}, this is crucial even at leading order. Actually, with only one term in the Hamiltonian, no generalized melon respecting the $\text{U}(N)^{2}$ symmetry could be built.

We introduce the Euclidean two-point function at finite temperature
\be\label{Gdef2} G(t) = \frac{1}{N}\bigl\langle\tr\psi_{\mu_{1}\cdots\mu_{r}}(t)\psi^{\dagger}_{\mu_{1}\cdots\mu_{r}}\bigr\rangle_{\beta}\, .\ee
It is real, satisfies $G(t+\beta)=-G(t)$, but it is not odd as in the Majorana case, except when $m=0$. In terms of the Fourier transform defined as in \eqref{GFourier} and the self-energy 
\be\label{Sigmadef2} \frac{1}{G_{k}}=m-\frac{2i\pi k}{\beta} + \Sigma_{k}\, ,\quad \Sigma(t) = \frac{1}{\beta}\sum_{k\in\mathbb Z+\frac{1}{2}}\Sigma_{k}\,e^{-\frac{2i\pi k t}{\beta}}\, ,\ee
the Schwinger-Dyson equation reads
\be\label{SDDirac} \Sigma (t) = 
\begin{cases} 16 |\la|^{2} G(t)^{2}G(-t) & \text{if $r=1$,}\\
(-1)^{\frac{r+3}{2}}(r+3)|\la|^{2}G(t)^{\frac{r+3}{2}}G(-t)^{\frac{r+1}{2}} & \text{if $r\geq 3$, $r+2$ prime.}
\end{cases}\ee
Following \cite{AFS}, we expect to find a rich physics for this model in the $(T,m)$-plane, with small and large black hole phases, a line of first order phase transition between them, terminating at a non-trivial critical point.

\section{\label{concSec}Open problems}

We conclude this paper with a list of interesting open problems.

An obvious generalization to consider is to extend the notion of index introduced in Sec.\ \ref{indsec} to graphs with external legs, hence boundaries. This should be straightforward, taking into account the fact that the $\gB_{(0ij)}$ can be surfaces with boundaries. It is expected that the index should decrease with the number of external legs, the number of connected components and the degree of the associated boundary graph. 

Another interesting task would be to characterize the full set of interaction bubbles for which our new scaling is optimal. More generally, finding the optimal scaling for an even wider class of interactions, in the line of \cite{Bonzometal}, is a non-trivial open problem. For example, one would like to identify the optimal scaling for the complete interaction of odd non-prime rank as well as for the melo-complete interactions of App.\ \ref{appendixB}. Even more ambitious would be to obtain the full classification of the leading graphs for as many interaction terms as possible, as we have done in Sec.\ \ref{applicationsSec} for the prime-complete interaction.

Even more difficult is the study of graphs beyond leading order. As briefly explained in Sec.\ \ref{MajoranaSec}, for interesting matrix-tensor models, the large $N$ and large $D$ limits cannot be obtained straightforwardly from the auxiliary field method, which makes the analysis of the subleading graphs very hard indeed. This is unlike the case of the SYK model and goes a long way in explaining why matrix-tensor (or purely tensor) models and SYK-like models will differ at subleading order, as studied explicitly for example in \cite{Bonzom:2017pqs}. This result is particularly significant because, unlike SYK-like models, matrix-tensor models are genuine quantum mechanical models to all orders.

Another open avenue is to define a field theoretic version of matrix-tensors and SYK models by breaking the tensor symmetry at the propagator level. This should allow one to distinguish ultraviolet from infrared modes, hence to launch a 
corresponding renormalization group analysis in the spirit of \cite{Geloun:2013saa,Rivasseau:2014ima}.

Our results and most of the literature on tensor models rely on the fact that each index of the tensors is associated with a symmetry group and thus a color. This property is lost if some symmetry properties on the indices are imposed. The first instance of this type was considered in \cite{ferra1} for matrix-tensor models based on Hermitian matrices, for which the $\text{U}(N)^{2}$ symmetry associated with the two indices of the matrices is broken down to a single $\text{U}(N)$. An  argument given in \cite{ferra1} shows that, in spite of this reduced symmetry, the large $D$ limit is still consistent at the planar level. The argument of \cite{ferra1} can be immediately generalized to the more general matrix-tensor models studied in the present paper: the large $D$ limit of matrix-tensor models for which the $\text{O}(N)^{2}\times\text{O}(D)^{r}$ symmetry is broken down to $\text{O}(N)\times\text{O}(D)^{r}$ by imposing the symmetry or antisymmetry of the associated real matrices still makes sense at the planar level. An extremely interesting open problem would be to generalize this result beyond the planar limit. As explained in \cite{ferra2}, the consistency of the large $D$ limit is lost at genus $g\geq 1$ if no further constraint is imposed on the Hermitian (or symmetric, or antisymmetric) matrices. The problematic graphs displayed in \cite{ferra2} disappear in the case of \emph{traceless} matrices and it is very natural to conjecture that the large $D$ limit is then consistent at all genera. The fact that the tracelessness condition could be sufficient is further supported by the numerical analysis presented in \cite{Klebanov:2017nlk} as well as by the arguments in  \cite{Gurau:2017qya}, which studies a particular bipartite model with two real symmetric tensors. Overall, it seems likely that well-defined large $D$ expansions can be defined for matrix-tensor models (or well-defined large $N$ expansions for pure tensor models) even when symmetry properties are imposed on the indices of the matrix-tensors. A full understanding of these expansions will require non-trivial extensions of the techniques presented in our paper and is an exciting research avenue for the future.

The physics of the quantum mechanical or field theoretical versions of matrix-tensor models in the large $N$ and large $D$ limits remains, to a large extent, to be uncovered. Recently, a non-trivial and unexpected structure for the phase diagram of models based on Dirac fermions was found \cite{AFS}. Our results, in particular the possibility to study the large $r$ limit of the model presented in Sec.\ \ref{DiracSec}, opens the way to a better analytical understanding of the phase transition and the non-trivial critical point discussed in \cite{AFS}. More generally, matrix-tensor models are known to capture basic qualitative properties associated with quantum black holes (quasi-normal behaviour, maximal chaos, macroscopic zero-temperature entropy, etc.), but a detailed and satisfactory picture of the relationship with black holes has not emerged yet. In particular, a model with a genuine gravity-like holographic dual has not been constructed and it is unclear how the black hole geometry can be seen directly from the quantum models. These are clearly outstanding research directions for the future. 

Finally, recall that random tensors were introduced originally in quantum gravity for a completely different reason, in order to sum over discrete geometries weighted by an analog of the Einstein-Hilbert action. Even though the relationship with the holographic point of view remains at present totally mysterious, it is fascinating to ponder on the possibility that a deep connection may, after all, exist.

\subsection*{Acknowledgments}

We would like to thank Tatsuo Azeyanagi, Nicolas Delporte, Paolo Gregori and Luca Lionni for useful discussions.
This research is supported in part by the Belgian Fonds National de la Recherche Scientifique FNRS (convention IISN 4.4503.15) and the F\'ed\'eration Wallonie-Bruxelles (Advanced ARC project ``Holography, Gauge Theories and Quantum Gravity''). G.V. is a Research Fellow at the Belgian F.R.S.-FNRS.

\appendix
\appendixpage
%

%
\section{A few remarks on MST interactions} \label{appendixA}

\subsection{\label{primebip}The complete bipartite interaction}

\begin{figure}
\begin{center}
{\includegraphics[width=6in]{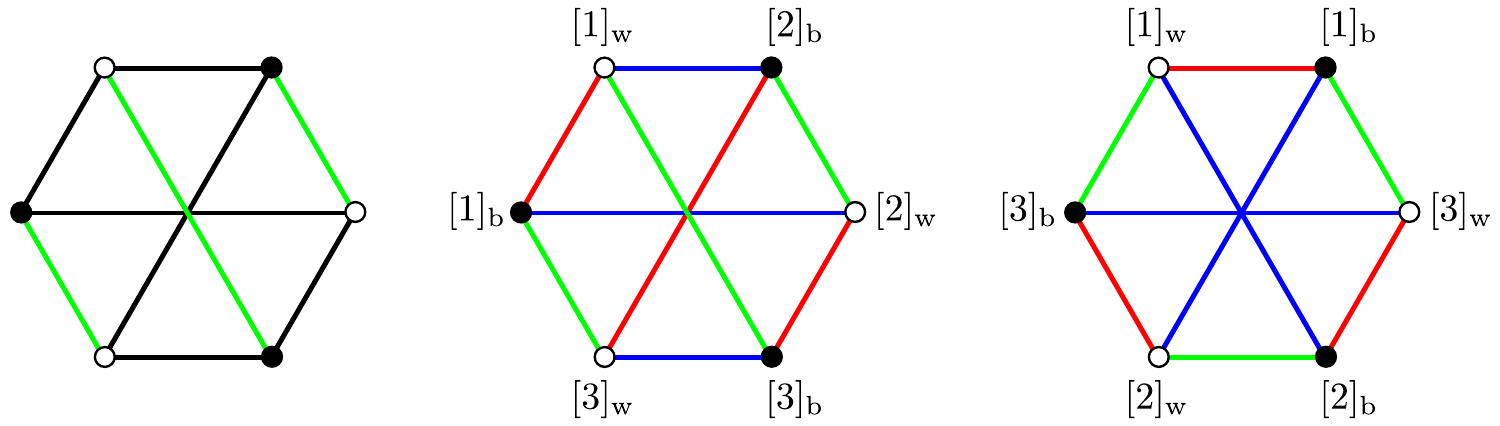}}
\end{center}
\caption{Edge-coloring for the complete bipartite graph $K_{3,3}$. Left: rule for the coloring of the edges of a given color, here green. Center: full edge-coloring and vertex labeling, here $1=\text{green}$, $2=\text{red}$ and $3=\text{blue}$. Right: equivalent (green,red)-polygonal representation; this type of representation is natural when $R$ is prime.}
\label{figureK33}
\end{figure}

The complete bipartite graph $K_{R,S}$ with $R$ black vertices and $S$ white vertices is edge-colorable with $\max{(R,S)}$ colors. In particular, $K_{R,R}$ is edge-colorable with $R$ colors. An explicit $R$-regular edge-coloring of $K_{R,R}$ is provided as follows. We first arrange the $2R$ vertices in the shape of a $2R$-sided polygon with alternating black and white vertices. We then label the black and white vertices cyclically as $[n]_{\text b}$ and $[n]_{\text w}$, where $n\in {\mathbb Z } / R {\mathbb Z }$, such that $[n]_\text{b}$ and $[n]_\text{w}$ follow each other. The vertices $[n]_\text{b}$ and $[m]_\text{w}$ are connected by an edge of color $i$, $1\leq i \leq R$, if and only if $n+m \equiv i$ where $\equiv$ denotes equality modulo $R$ as usual. The $R$-bubble obtained in this way is denoted by $\gK_{R,R}$. The above construction is illustrated for $\gK_{3,3}$ in Fig.\ \ref{figureK33}, which also contains a convenient $(ij)$-polygonal representation.

We have the following result.
\begin{lemma} \label{primebiplemma}
The $R$-bubble $\gK_{R,R}$ is maximally single-trace (MST) if and only if $R$ is a prime number.
\end{lemma}

\proof
The proof follows the same strategy as for Proposition \ref{prime}. Consider a face of colors $i$ and $j$, $1 \le i <j \le R$, whose length is denoted by $2q$. Using the edge-coloring explained above, we can write the $(ij)$-face starting from the vertex $[1]_\text{b}$ with the edge of color $i$ as $[1]_\text{b}[k_1\equiv i-1]_\text{w}[k_2]_\text{b} \cdots [k_{2q-1}\equiv j-1]_\text{w}[1]_\text{b}$. One can check inductively that $k_{2p}\equiv p(j-i)+1$ and $k_{2p+1}\equiv (p+1)i - pj - 1$ for $1\leq p\leq q-1$. As a result, $k_{2q-1}\equiv j-1$ is equivalent to $q(i-j)\equiv 0$. 

If $R$ is a prime number, this implies $q\equiv 0$ because $i-j\not\equiv 0$. The smallest possible solution is given by $q=R$. Therefore, the $(ij)$-face has length $2R$ and it visits all the vertices in $\gK_{R,R}$; hence the bubble is MST.

If $R$ is not prime, write $R=R_1R_2$, where $R_1$ and $R_2$ are integers with $1<R_1<R$ and $1<R_2<R$, and set $i-j=R_2$. The smallest possible solution to $q(i-j)\equiv 0$ is then $q=R_1<R$. This implies that the $(ij)$-face has length $2R_1<2R$ and therefore there are vertices in $\gK_{R,R}$ that are not visited by this face. As a result, the bubble is not MST. \qed

Note that, by Proposition \ref{MSToptprop}, our enhanced scaling is thus optimal for the bubbles $\gK_{R,R}$ when $R$ is prime. In the case of $\gK_{3,3}$, this was already noticed in \cite{LionniTh}, together with the full characterization of the leading graphs.

\subsection{Building MSTs from MSTs}

In this section, we further study the family of MST interactions. As explained previously, tensor models that contain only MST interactions are of special interest in our construction because the index of a vacuum Feynman graph is then given elegantly by Eq.\ \eqref{singfaceind}; see also Sec.\ \ref{MSTSec2}.
 
\begin{figure}
\begin{center}
{\includegraphics[width=6in]{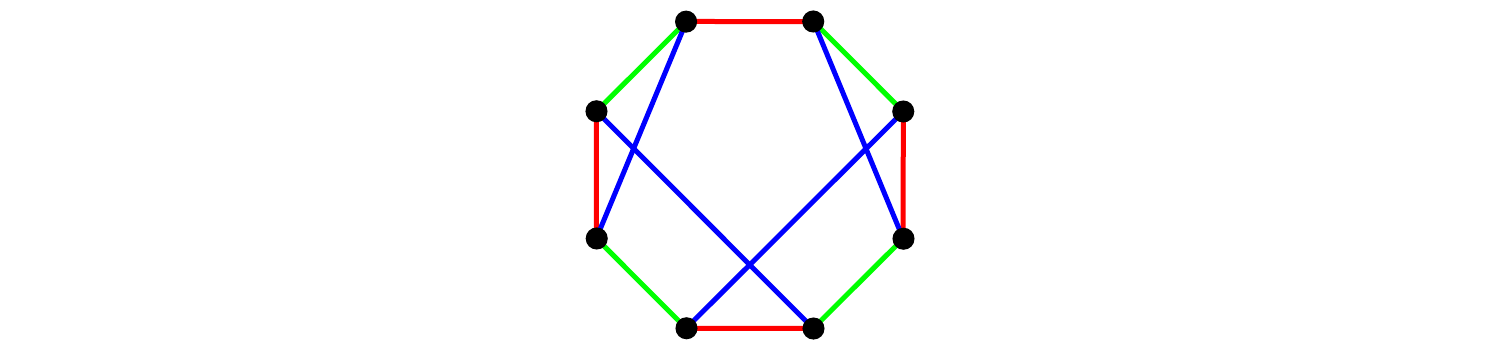}}
\end{center}
\caption{Example of an MST interaction that corresponds to a three-regular edge-colored graph with eight vertices.}
\label{figureMSTEx}
\end{figure}

Interesting examples of MST interactions include the complete interaction $\gK_{R+1}$ (see Fig.\ \ref{figureA}) for $R$ a prime number, studied in full detail in Sec.\ \ref{applicationsSec}, and the complete bipartite interaction $\gK_{R,R}$ (see Fig.\ \ref{figureK33}) for $R$ a prime number, presented in App.\ \ref{primebip}. Of course, there are many more possibilities, see Fig.\ \ref{figureMSTEx} for an example. To the best of our knowledge, a full classification of MST interactions is not known, but we gather a few remarks here.

No melonic graphs at rank greater than or equal to three with more than two vertices can be MST because melonic graphs have at least one face of length two \cite{melons}. 

Given two disjoint MST $R$-bubbles $\gB_1$ and $\gB_2$, each with $V \ge 3$ 
vertices, and two distinguished vertices $[\nu_1]$ in $\gB_1$ and $[\nu_2]$ in $\gB_2$, we can build another MST $R$-bubble $\gB_1 \cup_{[\nu_1][\nu_2]} \gB_2$ by removing the vertices $[\nu_{1}]$ and $[\nu_{2}]$ and gluing together the edges incident to $[\nu_1]$ with those incident to $[\nu_2]$, respecting the colors.

Conversely, if there exists a $R$-reducible cut in an MST $R$-bubble $\gB$, that is, a set of $R$ edges of different colors whose removal partitions the bubble into two disjoint connected components, we can perform the reverse operation, 
that is, cut the set of $R$ edges and glue vertices $[\nu_1]$ and $[\nu_2]$ at both ends of the cut. This rewrites $\gB$ as $\gB_1 \cup_{[\nu_1][\nu_2]} \gB_2$.

Any MST $R$-bubble with no such $R$-reducible cut is said to be irreducible. 
The prime-complete interaction $\gK_{R+1}$ and the prime-complete bipartite interaction $\gK_{R,R}$ are both irreducible.

\begin{figure}
\begin{center}
{\includegraphics[width=6in]{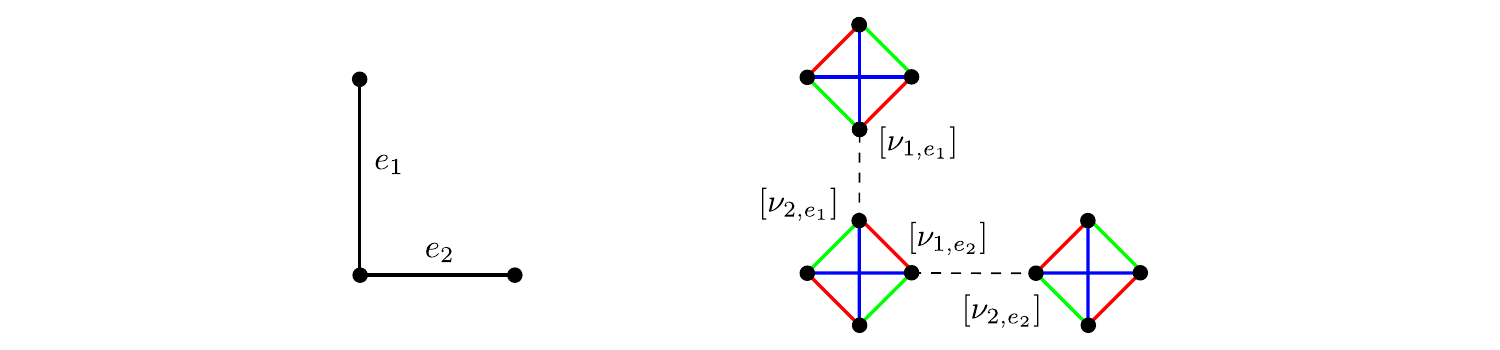}}
\end{center}
\caption{Example of a tetraedric colored rosette. Left: tree associated with the tetraedric colored rosette. Right: gluing of complete interactions $\gK_4$ according to the tree pattern. The resulting tetraedric colored rosette is the MST bubble depicted in Fig.\ \ref{figureMSTEx}.}
\label{figureTreeMST}
\end{figure}

Now, consider an abstract tree $\gT$ with $V$ vertices, associate to each vertex of $\gT$ 
an MST $R$-bubble and for each edge $e \in \gT$ a pair of vertices $[\nu_{1,e}]$ and $[\nu_{2,e}]$ in the MST bubbles at both ends of the edge such that the $2(V-1)$ vertices $[\nu_{1,e}]$, $[\nu_{2,e}]$ are all distinct. Then, we can glue together the $V$ MST bubbles according to the tree pattern, that is, we glue the edges incident to $[\nu_{1,e}]$ with those incident to $[\nu_{2,e}]$ for all $e \in \gT$, and get another MST $R$-bubble. For instance, if prime-complete interactions with $R+1$ vertices are glued together along a tree, we obtain a family we could call the ``$(R+1)$-edric colored rosettes.'' If $R=3$, we obtain ``tetraedric colored rosettes.'' This family is a kind of $\text{O}(N)$ non-bipartite tetraedric generalization of the melonic family. The construction is illustrated in Fig.\ \ref{figureTreeMST}.
\section{Melo-complete interactions} \label{appendixB}

In this appendix, we introduce a family of interaction bubbles that we call \emph{melo-complete}. At any rank, 
this family contains both the complete and the melonic bubbles, and ``interpolates" between these two extremes. 
Since this appendix requires the notion of boundary graph and is partly conjectural, we shall be sketchy.

A family that includes both the complete and the melonic bubbles appears to be required for a faithful higher-rank generalization of the Carrozza-Tanasa  model \cite{CarrozzaTanasa}. Recall indeed that this model, in addition to the tetraedric complete interaction $\gK_4$, also includes the quartic melonic interaction. 

First, let us recall the definition of boundaries in tensor models \cite{Bonzometal}. Consider a vacuum Feynman graph $\gB$ which corresponds, as explained in Sec.\ \ref{bubFeynmanSec}, to a $(R+1)$-bubble with the color $0$ associated with the propagators. Then, remove $k$ edges of color $0$ from it. The resulting graph, which we denote by $\gB_k$, corresponds to a $(R+1)$-colored graph with $2k$ free vertices, that is, vertices which do not have an incident edge of color $0$. Note that in this new graph $\gB_k$, there are as before faces of colors $0$ and $i$, that is, cycles made of edges of alternating colors $0$ and $i$. However, there are also paths made of edges of alternating colors $0$ and $i$ that do not close, because $\gB_k$ has free vertices. 

Any such $(R+1)$-colored graph $\gB_k$ with $2k$ free vertices defines naturally a $R$-bubble with $2k$ vertices, called its boundary bubble  and denoted by $\partial \gB_k$. It is obtained as follows. The set of vertices of $\partial \gB_k$ corresponds to the free vertices of $\gB_k$. In addition, there is an edge of color $i$ between the vertices $[\nu]$ and $[\nu']$ in $\partial \gB_k$ if and only if there is a path made of edges of alternating colors $0$ and $i$ between the free vertices $[\nu]$ and $[\nu']$ in $\gB_k$. In particular, all the faces in $\gB_k$ are forgotten. Note that in general $\partial \gB_k$ can have several connected components even when $\gB_k$ itself is connected.

Now, let us define the family of melo-complete interactions in the following way. Consider the $\text{O}(N)^R$ tensor model built around the complete interaction $\gK_{R+1}$ with optimal scaling. 
An $R$-bubble is called \emph{melo-complete} if it can be written as a connected component of the boundary bubble of 
some \emph{leading order} Feynman graph $\gB$ of that model after cutting an arbitrary number of edges of color $0$.

\begin{figure}
\centerline{\includegraphics[width=6in]{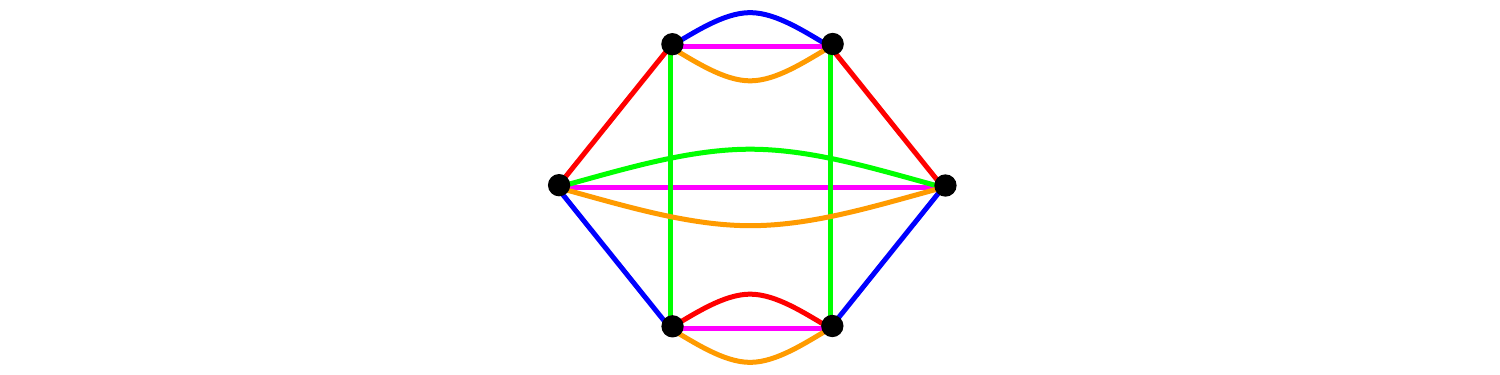}}
\caption{Example of a melo-complete bubble that corresponds to the boundary bubble of a generalized melon for the complete interaction $\gK_{6}$.}
\label{figureMeloComp}
\end{figure}

As an illustration, a melo-complete bubble which is neither complete nor melonic but ``in between'' is depicted in Fig.\ \ref{figureMeloComp}. It corresponds to a boundary bubble with six vertices obtained after cutting three edges of color $0$ in a generalized melon for the interaction bubble $\gK_{6}$.

We conjecture that like the melonic family, the melo-complete family is \emph{closed} under the operation of taking boundaries in the dominant sector.
More precisely

\medskip

\noindent{\bf 
Conjecture} {\it If $\gB_a$ is a melo-complete bubble and $\gB$ a Feynman graph of its leading sector under optimal scaling (assuming that it exists), then any connected component of a boundary bubble obtained from $\gB$ is again melo-complete.}

\medskip
It seems natural to define the generalized Carrozza-Tanasa model of rank $R$ by including in the action, together with the complete interaction, all melo-complete $R$-bubbles with $R+1$ or less vertices,
each with its own independent coupling constant $g_\gB$ and optimal scaling.  At rank $R=3$, the above prescription gives back the Carrozza-Tanasa action because in this case there are only two quartic melo-complete bubbles, the complete and the melonic bubbles.
%
%

%

%

%
\end{document}